\documentclass[aps,prd,twocolumn,showpacs,groupedaddress,superscriptaddress,nofootinbib,floatfix,preprintnumbers,longbibliography]{revtex4-1}

\usepackage[dvipdfmx]{graphicx}
\usepackage{dcolumn}
\usepackage{bm}

\usepackage{braket} 
\usepackage{color}
\usepackage{amsmath, amssymb}
\usepackage[colorlinks=true,citecolor=blue,linkcolor=red]{hyperref}
\usepackage{orcidlink}
\usepackage[normalem]{ulem}
\usepackage{quantikz}

\DeclareMathOperator{\Tr}{Tr}

\usepackage[caption=false]{subfig}

\begin{document}

\title{
Trapped-ion quantum simulation of (1+1)D SU(2) lattice\\gauge theory via loop-string-hadron formulation}

\author{Toshiaki Kaji\,\orcidlink{0000-0002-6532-7501}}
\email{kaji@icepp.s.u-tokyo.ac.jp}
\affiliation{
International Center for Elementary Particle Physics (ICEPP),
The University of Tokyo, 7-3-1 Hongo, Bunkyo-ku, Tokyo 113-0033, Japan.}

\author{Jesse R.~Stryker\,\orcidlink{0000-0002-4968-7988}}
\affiliation{
Physics Division, Lawrence Berkeley National Laboratory, Berkeley, CA 94720, USA.}
\affiliation{European Centre for Theoretical Studies in Nuclear Physics and Related Areas (ECT*), Fondazione Bruno Kessler, Strada delle
Tabarelle, 286, 38123 Trento (TN), Italy.}

\author{Yutaro Iiyama\,\orcidlink{0000-0002-8297-5930}}
\affiliation{
International Center for Elementary Particle Physics (ICEPP),
The University of Tokyo, 7-3-1 Hongo, Bunkyo-ku, Tokyo 113-0033, Japan.}

\author{Lento Nagano\,\orcidlink{0000-0003-4515-2115}}
\affiliation{Graduate School of Science and Technology, Keio University,\\
Yokohama, Kanagawa 223-8522, Japan}
\affiliation{
International Center for Elementary Particle Physics (ICEPP),
The University of Tokyo, 7-3-1 Hongo, Bunkyo-ku, Tokyo 113-0033, Japan.}

\author{Enrico Rinaldi\,\orcidlink{0000-0003-4134-809X}}
\affiliation{
Quantinuum, Partnership House, Carlisle Place, London SW1P 1BX, United Kingdom.}

\author{Christian W.~Bauer\,\orcidlink{0000-0001-9820-5810}}
\affiliation{
Physics Division, Lawrence Berkeley National Laboratory, Berkeley, CA 94720, USA.}
\affiliation{
Leinweber Institute for Theoretical Physics, Berkeley, CA 94720, USA.}

\author{Koji Terashi\,\orcidlink{0000-0001-6520-8070}}
\affiliation{
International Center for Elementary Particle Physics (ICEPP),
The University of Tokyo, 7-3-1 Hongo, Bunkyo-ku, Tokyo 113-0033, Japan.}

\date{\today}

\begin{abstract}
We present a quantum simulation of the SU(2) lattice gauge theory in 1+1 dimensions with one flavor of fermions, performed using the trapped-ion Quantinuum System Model H2.
This work is based on the loop-string-hadron (LSH) formulation of the gauge theory, and is distinguished by using qubits to represent both bosonic-gauge and fermionic-matter degrees of freedom.
Furthermore, we access the $j=1$ electric-flux sector beyond the $j_{\max}=1/2$ truncation commonly used in SU(2) quantum simulations.
To that end, we digitize an electric-field degree of freedom using a multiqubit register for representing the bosonic quantum number.
Our simulation therefore synthesizes several key ingredients relevant to quantum simulations in higher spatial dimensions, in particular explicit local gauge-field degrees of freedom represented here by a multiqubit bosonic register, while relying solely on circuit-construction optimizations that are expected to scale polynomially with lattice size.
This work demonstrates that modern quantum hardware is increasingly able to accommodate the features of complicated but physically important lattice gauge theories while successfully reproducing the expected Trotterized dynamics.
\end{abstract}

\maketitle

\section{Introduction}
One of the key applications for quantum computers has been identified as the simulation of lattice gauge theories \cite{Banuls:2019bmf,Bauer:2023qgm,Byrnes:2005qx,Zohar:2015hwa,Dalmonte:2016alw,Aidelsburger:2021mia,Zohar:2021nyc,Klco:2021lap,Bauer:2022hpo,DiMeglio:2023nsa,Halimeh:2023lid}.
Lattice gauge theory is the only available numerical tool to make predictions for interacting field theories, in particular those describing the interactions of elementary particles, without the need to rely on perturbation theory.
Most lattice gauge theory calculations are based on evaluating the path integral using Monte Carlo integration.
Since Monte Carlo integration requires a positive definite integrand, it is generally applied to theories defined in Euclidean spacetime in order to avoid the sign problem~\cite{deForcrand:2009zkb,Troyer:2004ge}. 
The requirement for a positive-definite integrand limits the kinds of problems that can be studied with Monte Carlo integration, and in particular does not allow to directly study the dynamics of the field theory. 
Recently, approaches based on the Hamiltonian formulation \cite{Kogut:1974ag} have attracted considerable attention, driven in large part by the ability to efficiently perform the required calculations using quantum computers. This is expanding the scope of theoretical investigations possible
\cite{Martinez:2016yna,Klco:2019evd,Ciavarella:2021nmj,Davoudi:2025rdv}
and in particular allowing physicists to study the dynamics of fundamental processes of nature~\cite{Jordan:2011ci,Bauer:2021gup,Davoudi:2024wyv,Bauer:2025nzf}.

Our current best understanding of nature at small scales, the Standard Model, has multiple layers of complexity.
The strong interaction sector, corresponding to the SU(3)-color gauge symmetry of the Standard Model, is especially complex because of its nonperturbative coupling at scales larger than approximately one femtometer. A proper representation of the physics of hadrons should account for the SU(3) gauge fields, multiple flavors of massive chiral fermions, and 3+1 dimensional spacetime.
When calculations are based on lattice field theory, one must also take limits such as zero lattice spacing (i.e., the continuum limit), infinite volume, and infinite Hilbert space dimension~\cite{Carena:2021ltu,Clemente:2022cka,Kane:2025ybw,Halimeh:2024bth}.
In practice, simulations cannot start with all this complexity built in.
Instead, toy models of increasing complexity are studied, such that more and more features of QCD or the Standard Model are increasingly understood in terms of their practical implementation.
The simplest toy model exhibiting several key features of more complicated gauge theories---QCD included---is the theory with a single flavor of fermion, the simpler gauge group of U(1), and only one spatial dimension. 
This is the Schwinger model, describing quantum electrodynamics in 1+1 dimensions~\cite{Schwinger:1962tp,Martinez:2016yna}.
In this work, we take one step in the direction of QCD by generalizing the gauge group of the Schwinger model from U(1) to SU(2).

Another important issue is the encoding of gauge degrees of freedom.
It is well known that gauge theories in a single spatial dimension and open boundary conditions do not contain dynamical gauge degrees of freedom. 
For quantum simulation of LGTs containing dynamical fermions, most works have therefore integrated out the gauge fields, which results in the fermion representation of the Hamiltonian
\cite{Martinez:2016yna,Atas:2021ext,Than:2024zaj,Ilcic:2026cac,Klco:2018kyo,Shaw:2020udc,Nguyen:2021hyk,Atas:2022dqm,Nagano:2023uaq,Farrell:2024fit}. 
Such ``purely fermionic'' Hamiltonians for 1+1 dimensions are attractive from the point of view of near-term quantum simulation because the degrees of freedom are easily mapped to qubits and the overall number of logical qubits required is relatively low.
Unfortunately, the total removal of gauge fields is not possible for LGTs in $d\ge 2$, where gauge degrees of freedom are dynamical and must be included in the theory~\cite{Haase:2020kaj,Paulson:2020zjd,Bauer:2021gek,Kan:2021xfc,Kane:2022ejm,Ciavarella:2024fzw,Balaji:2025afl,Hidalgo:2026zsz,Draper:2026bcj,Balaji:2025yua}.

While some previous works explicitly keep the gauge fields in 1+1 dimensions~\cite{Klco:2019evd}, they mostly rely on the smallest nontrivial truncation---e.g. $j\in \{0,1/2\}$ in the electric basis representation---which is not sufficient for performing calculations that can be used to extrapolate to the continuum limit~\cite{Hackett:2018cel,Tong:2021rfv,Ciavarella:2025tdl}.
In general, the bosonic degrees of freedom should be represented by multiqubit registers or qudits to better approximate the bosonic Hilbert space. Explorations using qudits have indeed been intensively studied recently~\cite{Meth:2023wzd,Calajo:2024qrc,Jiang:2025ufg,John:2026gut}.

A complementary approach is to represent these higher-dimensional degrees of freedom using multiqubit registers. This approach is particularly relevant
because manufacturers have generally expressed more interest in refining their qubits and taking advantage of the error-correction theory that is well understood for two-level systems than in developing higher-dimensional qudits for end users.
For the U(1) case, Refs.~\cite{Crippa:2024hso,Rosanowski:2025nck} encode the three electric-field states $e\in\{-1,0,1\}$ in a two-qubit register on trapped-ion hardware. Ref.~\cite{Mou:2025iiu} also encodes gauge fields into multiqubits in a different manner.
For $\mathrm{SU}(2)$, Refs.~\cite{Froland:2025bqf,Froland:2026aff} used a mixed-basis formulation, in which the local \emph{continuous} gauge degrees of freedom are digitized onto multiqubit registers (two qubits per each degree of freedom) for execution on quantum hardware.
Our focus is different from Refs.~\cite{Froland:2025bqf,Froland:2026aff} because we use the (naturally discrete) electric basis, and, within that context, we raise the electric truncation beyond the typical $j\leq 1/2$ cutoff by using a multiqubit encoding.
Our work is further distinguished from Refs.~\cite{Froland:2025bqf,Froland:2026aff} because our Hamiltonian retains manifest locality and translational invariance\footnote{There is no true translational symmetry for the open boundary conditions we consider. We mean to say only that the local operators defined on the lattice are related to each other by translations \emph{up to} boundary effects.}.

Considerations such as digitization, truncation, locality, and the types of degrees of freedom are all related to the choice of formulation and basis for a quantum field theory.
In recent years, several formulations of non-Abelian LGTs have been proposed for quantum simulation~\cite{Anishetty:2009nh,Mathur:2015wba,Lamm:2019bik,Alexandru:2019nsa,Davoudi:2020yln,Ciavarella:2021nmj,Ciavarella:2024fzw,Ciavarella:2025bsg}. As for the choice of basis, most works have focused on electric bases, in which the electric contribution to the gauge Hamiltonian is diagonal. As already discussed, these approaches require larger truncations as the coupling constant is lowered, as is the case for the continuum limit of lattice QCD. Alternative approaches that have better scaling as $g \to 0$ are also available~\cite{DAndrea:2023qnr,Grabowska:2024emw,Kaplan:2018vnj,Unmuth-Yockey:2018xak,Haase:2020kaj,Bender:2020ztu,Bauer:2021gek}, but will not be considered here.  
In this work, we consider a particular formulation using an electric basis called the loop-string-hadron (LSH) formulation \cite{Raychowdhury:2019iki,Raychowdhury:2018osk,Kadam:2022ipf,Kadam:2024ifg,Burbano:2024uvn}, which facilitates a natural truncation in the electric basis. Furthermore, it is constructed so that the nonabelian Gauss's law constraints are inbuilt, effectively trading them for simpler Abelian constraints that are referred to as the Abelian Gauss's law (AGL). 
While the theoretical formulation and circuit construction \cite{Davoudi:2022xmb} of the LSH-based theory have been available for some time, their simulation on quantum hardware has only begun to be explored recently \cite{Ilcic:2026cac}.
Ref.~\cite{Ilcic:2026cac} exploits the special structure of the 1+1-dimensional theory, in which the bosonic flux $n_l$ is determined by the boundary flux and fermionic configuration, to obtain a hardware-efficient purely fermionic encoding. Our focus is complementary: we retain the local bosonic degrees of freedom as part of the encoding, including a multiqubit bosonic register, in order to investigate a hardware setting that preserves such a key ingredient of theories beyond one spatial dimension.

In this paper, we study the one-dimensional SU(2) LGT in the presence of dynamical fermions using the LSH formalism.
Specifically, we identify the minimal setting (four-site lattice with specific open boundary conditions) that allows us to explore the states beyond $j=1/2$ and implement a Trotterized real-time evolution on commercial quantum hardware provided by Quantinuum based on this setting.
We use an adaptation of the circuit construction given in Ref.~\cite{Davoudi:2022xmb} to study the time evolution of the electric-field energy density throughout the lattice, starting from a state with nontrivial flux. 
To bridge the gap between the general construction in Ref.~\cite{Davoudi:2022xmb} and a hardware-ready circuit, we first analytically optimize the circuit using the specific boundary condition we identified. Second, the resulting circuit is processed with the transpiler provided by Quantinuum (pytket~\cite{Sivarajah_2020}). 
To improve the quality of the hardware results, we consider two techniques separately: one is a standard error mitigation technique (rescaling) and the other is a post-selection method.
The post-selection discards the states that do not satisfy (Abelian) Gauss's law constraints, which are easily checked thanks to the LSH formalism.
We show at the circuit level that AGL constraints can detect not only bit-flip ($X$) errors, but also certain phase-flip ($Z$) errors that are converted into detectable $X$- or $Y$-type errors by subsequent basis-changing operations.
Applied to the hardware data, the two mitigation methods reduce the overall root-mean-square deviation from the noiseless Trotterized-circuit baseline by approximately $60\%$ and $75\%$, respectively, relative to the raw hardware data, with post-selection achieving the smaller deviation at the cost of an increasing discard rate with circuit depth.
We expect this minimal demonstration to be an important step toward larger-scale simulations, since LGT Hamiltonians are local and should scale by composing local building blocks~\cite{Stryker:2021asy,Grabowska:2022uos,Kane:2022rlr}. 

The paper is structured as follows.
In Sec.~\ref{sec: background}, we review the setup of the (1+1)D SU(2) lattice gauge theory in the loop-string-hadron formulation.
In Sec.~\ref{sec: methods}, we explain the circuit construction involved with real-time evolution, relevant optimizations, and two types of error mitigation we employ.
In Sec.~\ref{sec: results}, we discuss the application of the error mitigation methods and present the data we gathered from the H2-2 platform.
We interpret our results in Sec.~\ref{sec: discussion} and conclude in Sec.~\ref{sec: conclusion}.

\section{\label{sec: background}Theory foundation in LSH formulation}
\subsection{LSH degrees of freedom and Hamiltonian} \label{subsec:model}
\begin{figure}[t!]
    \centering
    \includegraphics[width=0.99\linewidth]{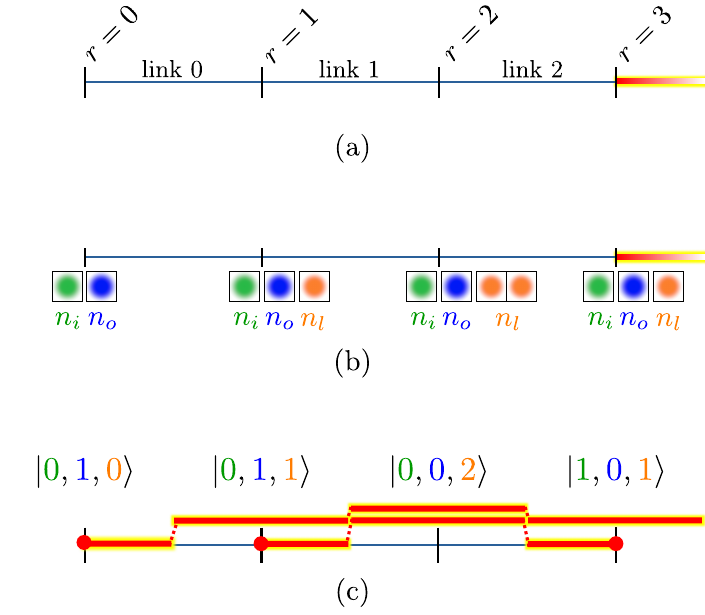}
    \caption{The lattice and its degrees of freedom. (a) A depiction of the lattice with four staggered sites, open boundary conditions, and a single unit of external flux sourced at the right-side boundary. (b) A depiction of the LSH degrees of freedom relevant to the system. Each box corresponds to one qubit used to represent the indicated quantum numbers. (c) A cartoon illustration of the LSH basis state $\ket{(0,1,0),(0,1,1),(0,0,2),(1,0,1)}$, with a certain set of specified boundary flux values. Red dots are drawn to illustrate the Abelian Gauss's law: the units of gauge flux at either end of a link must be balanced.}
    \label{fig: lattice dof}
\end{figure}

We consider a lattice of $N_s$ staggered sites and open boundary conditions.
Sites are labeled from left to right by $r\in \{0,1,\ldots,N_s-1\}$.
Each site $r$ is connected to site $r+1$ by a gauge link.
We make the conventional choice to label gauge links by the site to their left.
The links are therefore labeled by $r\in\{0,1,\ldots,N_s - 2 \}$.
Fig.~\ref{fig: lattice dof}(a) depicts the site and link labeling for the case of $N_s=4$.

In the LSH formulation, the elementary degrees of freedom are fields associated to each site $r$: two fermionic occupation numbers, $n_i(r)\in \{0,1\}$ and $n_o(r)\in\{0,1\}$, and one bosonic occupation number, $n_l(r)\in \{0,1,2,...\}$.
The above-mentioned quantum numbers characterize a basis that may be referred to interchangeably as the ``charge'', ``flux'', or ``electric'' eigenbasis.
A basis state of the lattice is a tensor-product state: $\otimes_{r=0}^{N_s-1} \ket{n_i(r),n_o(r),n_l(r)}$.
An example basis state of the lattice is depicted in Fig.~\ref{fig: lattice dof}.

To express various physical quantities and the Hamiltonian, there are a few operators to be aware of:
\begin{itemize}
    \item Number operators: $\hat{n}_i(r)$, $\hat{n}_o(r)$, and $\hat{n}_l(r)$.\\
    By definition, these operators are diagonal in the basis introduced above.
    One often encounters the derived number operators
\begin{align}
    \label{eq: N_L}
    \hat{N}^{L}(r) &\coloneq \hat{n}_{l}(r)+\hat{n}_{o}(r)(1-\hat{n}_{i}(r)),\\
    \label{eq: N_R}
    \hat{N}^{R}(r) &\coloneq \hat{n}_{l}(r+1)+\hat{n}_{i}(r+1)(1-\hat{n}_{o}(r+1)),
\end{align}
where $\hat{N}^L(r)$ [$\hat{N}^R(r)$] counts the number of flux units on the left (right) side of link $r$.\footnote{Some papers use a different convention, in which the flux count at the right side of the link joining sites $r$ and $r+1$ is called ``$\hat{N}_R(r+1)$'' instead of $\hat{N}_R(r)$.}
Note that the flux number operators in Eqs.~\eqref{eq: N_L}-\eqref{eq: N_R} correspond to two times the SU(2)-irrep quantum number $j$ of the Kogut-Susskind formulation~\cite{Kogut:1974ag}.
    \item Fermionic ladder operators: $\hat{\chi}_i$, $\hat{\chi}_i^\dagger$, $\hat{\chi}_o$, and  $\hat{\chi}_o^\dagger$.\\
    $\hat{\chi}_i(r)$ [$\hat{\chi}_i^\dagger(r)$] is the annihilation (creation) operator associated with $n_i(r)$, and similarly for $i\to o$. The fermionic creation and annihilation operators satisfy the standard commutation and anticommutation relations one would expect for fermionic fields on a lattice \cite{Raychowdhury:2019iki,Davoudi:2022xmb}.
    \item Bosonic, normalized ladder operators: $\hat{\Gamma}(r)$ and $\hat{\Gamma}^\dagger(r)$.\\
    $\hat{\Gamma}(r)$ is a lowering operator for $n_l(r)$: 
    \begin{align}
     &\quad\hat{\Gamma}(r)\ket{n_i(r),n_o(r),n_l(r)}\notag
     \\&
     =(1-\delta_{n_l(r),0})\ket{n_i(r),n_o(r),n_l(r)-1} \,,  
    \end{align}
    and its Hermitian adjoint is a raising operator satisfying
\begin{align}
&\quad\hat{\Gamma}^\dagger(r)\ket{n_i(r),n_o(r),n_l(r)}\notag
\\&=\ket{n_i(r),n_o(r),n_l(r)+1}   \,.
\end{align}
These bosonic, normalized ladder operators above are not standardly defined, i.e., they do not obey conventional bosonic commutation relations, however it is true that the bosonic ladder operators at site $r$ commute with all operators at any other site $r'\neq r$, as one would expect for bosons.
\end{itemize}

Using the basic operators introduced above, the Hamiltonian is given by~\cite{Davoudi:2022xmb}
\begin{align}
    H = H_M + H_E + H_I\,,
\end{align}
where the three distinct contributions are defined in the LSH formulation as follows. The mass energy, $H_M$, is a function of the LSH fermionic occupation numbers:
\begin{align}
H_{M}&=\mu\sum_{r=0}^{N_s-1}(-1)^{r}
\left(
\hat{n}_{i}(r)+\hat{n}_{o}(r)
\right),
\label{eq:mass-term-LSH}
\end{align}
with $\mu$ being a mass parameter proportional to the bare mass.
The electric energy, $H_E$, is a function of the number of elementary flux units along the links.
In the LSH formulation, there is some arbitrariness in how $H_E$ is defined; we use the choice
\begin{align}
    H_{E}&=\sum_{r=0}^{N_s-2}
\frac{\hat{N}^{L}(r)}{2}\left(
\frac{\hat{N}^{L}(r)}{2}+1\right). \label{eq:electric-term-LSH}
\end{align}

The most complicated part of the Hamiltonian is the interaction energy, $H_I$.
The hopping terms in $H_I$ are expressed using conditional ladder operators, e.g., $\hat{\Gamma}^\dagger(r)^{\hat{n}_i(r)}$, which is locally defined at site $r$ by
\begin{align}
    \hat{\Gamma}^{\dagger \, \hat{n}_i} \ket{n_i,n_o,n_l} &=
    \begin{cases}
        \ket{n_i,n_o,n_l}, \quad \, \mathrm{if}\ n_i=0, \\
        \hat{\Gamma}^\dagger \ket{n_i,n_o,n_l}, \, \mathrm{if} \ n_i=1.
    \end{cases}
\end{align}
The expression for $H_I$ is then
\begin{widetext}
\begin{align}
H_{I}&=
x\sum_{r=0}^{N_s-2}
\Big[
\left(\hat{\chi}^{\dag}_{i}(r+1)\hat{\Gamma}^{\dag}(r+1)^{\hat{n}_{o}(r+1)}\right)
\left(\hat{\chi}_{i}(r)\hat{\Gamma}^{\dag}(r)^{1-\hat{n}_{o}(r)}\right) 
\sqrt{
\frac{\hat{n}_{l}(r)+1+\hat{n}_{o}(r)}{\hat{n}_{l}(r)+1+\hat{n}_{o}(r+1)}
} \notag
\\ 
&\qquad+
\left(\hat{\chi}^{\dag}_{o}(r)\hat{\Gamma}^{\dag}(r)^{\hat{n}_{i}(r)}\right)
\left(\hat{\chi}_{o}(r+1)\hat{\Gamma}^{\dag}(r+1)^{1-\hat{n}_{i}(r+1)}\right)
\sqrt{
\frac{\hat{n}_{l}(r+1)+1+\hat{n}_{i}(r+1)}{\hat{n}_{l}(r+1)+1+\hat{n}_{i}(r)}}
+\text{H.c.}\Big] ,
\label{eq:interaction-term-LSH}
\end{align}
\end{widetext}
with $x$ being a dimensionless parameter.
The form in Eq.~\eqref{eq:interaction-term-LSH} is not identical to the original expression in Ref.~\cite{Raychowdhury:2019iki}, but is equivalent, as explained in Ref.~\cite{Davoudi:2022xmb}.
Full details about the operators, their properties, and the Hamiltonian are in Ref.~\cite{Raychowdhury:2019iki}.

In the present work, we always consider $N_s=4$ lattice sites and external boundary flux equal to zero (one) elementary units to the left (right) of site $r=0$ ($r=N_s-1$).
Fig.~\ref{fig: lattice dof}(a) depicts the four-staggered-site lattice, its three gauge links, and the boundary-flux configuration.
Fig.~\ref{fig: lattice dof}(b) illustrates the mapping of local LSH quantum numbers $n_i(r)$, $n_o(r)$, and $n_l(r)$ to qubits, where each box represents one qubit.
In the most general setting, the $n_l(r)$ quantum numbers are formally unbounded, however the boundary conditions and AGL constraints restrict what values of $n_l(r)$ are physically observable.
At site $r=0$, $n_l$ is fixed to zero; at sites $r=1$ and $r=3$, one has $0\leq n_l\leq1$; and at site $r=2$, one has $0\leq n_l \leq 2$.
The three-dimensional Hilbert space of $n_l(r=2)$ is mapped to the first three computational basis states of a two-qubit register ($\ket{00}$, $\ket{01}$, and $\ket{10}$).

In the following, we summarize the qubit-operator decompositions on which the quantum circuit is based.
For the LSH quantum numbers, we use the mappings
\begin{align}
    \label{eq: n_i expression}
    \hat{n}_i(r) &= (1-Z_{i}(r))/2, \\
    \hat{n}_o(r) &= (1-Z_{o}(r))/2, \\
    \label{eq: n_l expression}
    \hat{n}_l(r) &= \sum_{k=0} 2^{k-1} (1-Z_{l,k}(r)),
\end{align}
where $Z_i(r)$ [$Z_o(r)$] is the Pauli-$Z$ operator associated with the qubit for $n_i(r)$ [$n_o(r)$], and $Z_{l,k}(r)$ is that associated with the $k$-th qubit used in the binary representation of $n_l(r)$, with $k=0$ denoting the least-significant bit.
Then the mass terms in Eq.~\eqref{eq:mass-term-LSH} are replaced by
\begin{align}
H_{M}(r)
&=\sum_{j=1}^{2}H_{M}^{(j)}(r)\,,
\\
H_{M}^{(1)}(r) &= \frac{\mu}{2}(-1)^{r+1}Z_{i}(r)\,,
\\
H_{M}^{(2)}(r) &= \frac{\mu}{2}(-1)^{r+1}Z_{o}(r)\,.
\end{align}
The electric terms are decomposed into three summands as
\begin{align}
H_{E}(r)&=H_{E}^{(1)}(r) + H_{E}^{(2)}(r) + H_{E}^{(3)}(r),
\\
H_{E}^{(1)}(r) & = \frac{1}{2}\hat{n}_{l}(r)\,,
\\
H_{E}^{(2)}(r) & = \frac{1}{4}\hat{n}_l(r)^{2}\,,
\\
H_{E}^{(3)}(r) & = \frac{1}{2}\left[\hat{n}_l(r)+\frac{3}{2}\right] \hat{n}_{o}(r) \left(1-\hat{n}_{i}(r)\right)\,.
\end{align}
The expression of each $H_E^{(i)}(r)$ in terms of qubit operators follows from substituting Eqs.~\eqref{eq: n_i expression}-\eqref{eq: n_l expression}.

\pagebreak
The interaction energy is partitioned into $2(N_{s}-1)$ summands, i.e.,
\begin{align}
    H_I &= \sum_{r=0}^{N_{s}-2} \sum_{j=1}^{2}H_I^{(j)}(r) ,
\end{align}
and we use the same Jordan-Wigner transformation as Ref.~\cite{Davoudi:2022xmb}.
Following Ref.~\cite{Davoudi:2022xmb}, we diagonalize each hopping subterm using a singular-value-decomposition (SVD)-based construction. 
Rather than writing out the $H_I^{(j)}(r)$ expressions explicitly, we give their SVD-diagonalized forms:
\begin{widetext}
\begin{align}
    &\mathcal{U}_{\text{SVD}}^{(1)} (r) H_{I}^{(1)}(r)
\mathcal{U}_{{\text{SVD}}}^{(1) \, \dag} (r) \nonumber \\
    &=x\ket{1}\bra{1}_{i,r}Z_{i,r+1}Z_{o,r}(1-\ket{1}\bra{1}_{o,r+1}\ket{\Lambda}\bra{\Lambda}_{l,r})
(1-\ket{0}\bra{0}_{o,r}\ket{\Lambda}\bra{\Lambda}_{l,r})
\mathcal{D}(\hat{n}_{l}(r),\hat{n}_{o}(r),\hat{n}_{o}(r+1))\,, \label{eq:uhu} \\ 
    &\mathcal{U}_{\text{SVD}}^{(2)} (r) H_{I}^{(2)}(r)
\mathcal{U}_{{\text{SVD}}}^{(2) \, \dag} (r) \nonumber \\&=x\ket{1}\bra{1}_{o,r+1}Z_{o,r}Z_{i,r+1}(1-\ket{1}\bra{1}_{i,r}\ket{\Lambda}\bra{\Lambda}_{l,r+1})
(1-\ket{0}\bra{0}_{i,r+1}\ket{\Lambda}\bra{\Lambda}_{l,r+1})
\mathcal{D}(\hat{n}_{l}(r+1),\hat{n}_{i}(r+1),\hat{n}_{i}(r))\,, \label{eq:uhu2} \\
    \label{eq: svd-unitary 1}
    \mathcal{U}_{\text{SVD}}^{(1)} &\coloneq
    \mathsf{H}_{i,r+1} \left(\ket{0}\bra{0}_{i,r+1} + \ket{1}\bra{1}_{i,r+1} X_{i,r}  (\lambda^{-}_{l,r+1})^{\hat{n}_{o,r+1}} (\lambda^{-}_{l,r})^{1-\hat{n}_{o,r}} \right) , \\
    \label{eq: svd-unitary 2}
    \mathcal{U}_{\text{SVD}}^{(2)} &\coloneq
    \mathsf{H}_{o,r} \left(\ket{0}\bra{0}_{o,r} + \ket{1}\bra{1}_{o,r} X_{o,r+1}  (\lambda^{-}_{l,r})^{\hat{n}_{i,r}} (\lambda^{-}_{l,r+1})^{1-\hat{n}_{i,r+1}} \right) ,
\end{align}
\end{widetext}
with $\mathsf{H}= \frac{1}{\sqrt{2}}\left( \begin{array}{cc}
    1 & 1 \\
    1 & -1
\end{array} \right) $ being the qubit Hadamard gate,
\begin{align}
\mathcal{D}(p,n,n')
&\coloneq
\sqrt{\frac{p+1+n}{p+1+n'}},
\end{align}
and $\lambda^{-}$ denoting a cyclic lowering operator acting on an integer bosonic register.
$\Lambda$ is the maximum value of $n_l$, so representing $n_l\in\{0,\ldots,\Lambda\}$ in binary requires $\lceil\log_2(\Lambda+1)\rceil$ qubits.
In the following, we refer to the operators on the right-hand sides of Eqs.~\eqref{eq:uhu} and~\eqref{eq:uhu2}, which are diagonal in the computational basis, as the diagonal parts (or diagonal blocks) of the corresponding hopping subterms.

Eqs.~\eqref{eq:uhu}–\eqref{eq: svd-unitary 2} display the SVD construction with $y'$ as the control qubit in the $p$-based diagonal form used explicitly in Ref.~\cite{Davoudi:2022xmb}. In the hardware circuits, we additionally exploit a complementary SVD construction with $x'$ as the control qubit and the AGL-related freedom to represent the constrained diagonal operator using either bosonic register. These equivalent diagonal forms and their boundary-condition-dependent circuit costs are discussed in Appendix~\ref{app:preoptimization}.

\subsection{Symmetries and constraints}
The Hamiltonian exhibits a few symmetries.
There are two global U(1) charges, $\sum_r n_i(r)$ and $\sum_r n_o(r)$. Sometimes these are instead given in the linear combinations $\sum_r (n_i(r) +  n_o(r))$ and $\sum_r (n_i(r) -  n_o(r))$ \cite{Davoudi:2020yln}; the former counts the total fermion number while the latter indicates how many elementary units of electric flux are sourced by or sunk by the system as one traverses the entire system from left to right.
Finally, there are local Abelian gauge constraints, the Abelian Gauss's laws (AGLs), that require the conservation of electric flux units as one traverses from one end of a gauge link to the other; the AGL constraints are $N_s-1$ in number.

It is also true that the left- and right-boundary fluxes are constants of motion, denoted by $\hat{N}^R(-1)$ and $\hat{N}^L(N_s-1)$, respectively.
Together with the AGL constraints, the conserved boundary fluxes confine the dynamics to the subspace specified in Sec.~\ref{subsec:model}. 
For the boundary conditions considered here, the site-dependent ranges of $n_l(r)$ given above therefore remain valid throughout the ideal time evolution; in particular, $n_l(r=2)\in\{0,1,2\}$, so a two-qubit bosonic register at $r=2$ is sufficient to represent the full physically accessible range.

\section{\label{sec: methods}Methods}
\subsection{Observables and initial state}
To study the performance of the quantum simulation, we aim to use as observables the electric-field energy profile.
This is expressed using the local electric terms at each link, which are defined by 
\begin{align}
    H_E(r)=
\frac{\hat{N}^{L}(r)}{2}\left(
\frac{\hat{N}^{L}(r)}{2}+1\right)\,. \label{eq:obs}
\end{align}
This is a useful probe to explore interesting physics, such as string breaking~\cite{Hebenstreit:2013baa,Surace:2019dtp,Magnifico:2019kyj,De:2024smi,Cochran:2024rwe,Gonzalez-Cuadra:2024xul,Joshi:2026hfe,Xu:2026ibi}. At the same time, it can be easily measured since it has a diagonal form in the computational basis.

To pair with this choice of observable, we opt to use an initial state with large electric energy.
The initial state is defined as 
\begin{align}
    \otimes_r \ket{n_i(r),n_o(r), n_l(r)}
    =
    \ket{0,1,0}\ket{0,1,1}\ket{0,0,2}\ket{1,0,1},
\end{align}
the same state depicted in Fig.~\ref{fig: lattice dof}(c).
This choice of initial state with high electric energy is deliberate because it contains $n_l(2)=2$, hence it directly accesses the $j=1$ subspace of the Hilbert space while providing abundant energy to be monitored as it redistributes and changes forms during the subsequent time evolution. One can see that the initial state belongs to the matter-superselection sector
\begin{align}
    \label{eq: matter superselection sector}
    \biggl(\sum_r n_i(r),\sum_r n_o (r) \biggr) & =(1,2) .
\end{align}

\subsection{Circuit structure and optimization}
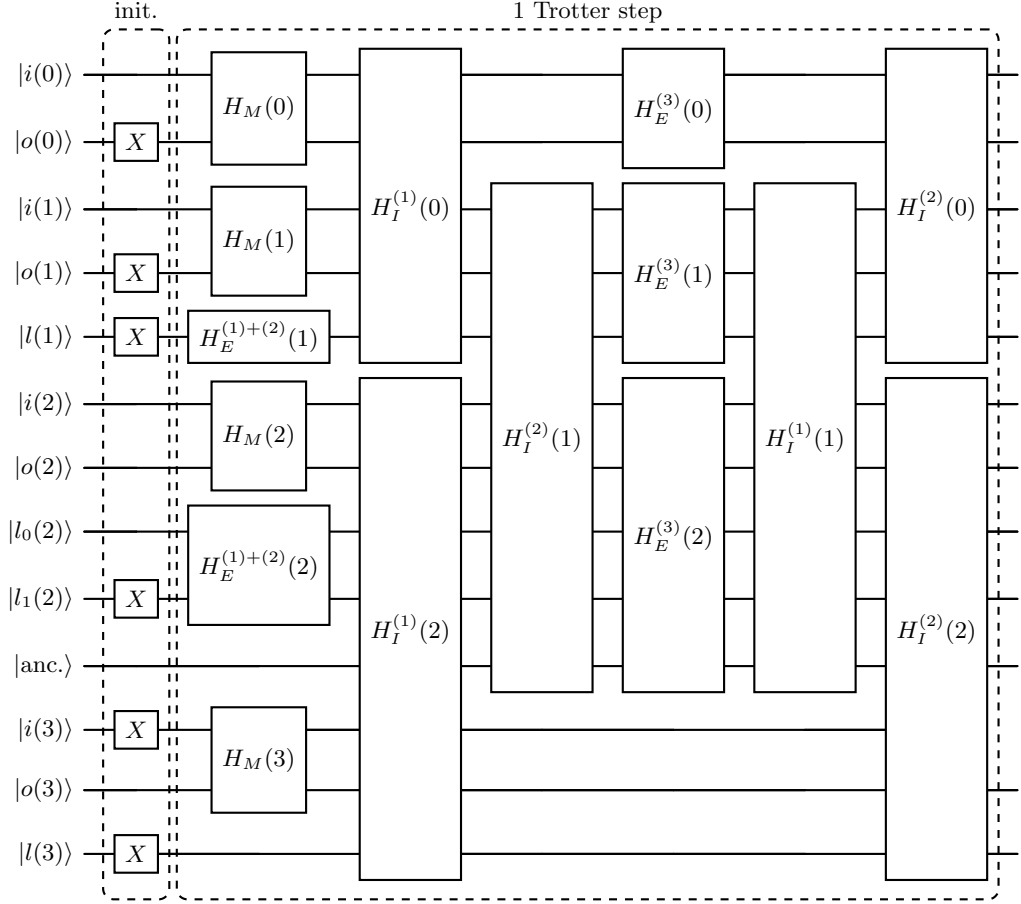
\begin{figure*}[t]
    \centering
    \begin{quantikz}[row sep=0.2cm, column sep=0.4cm]
        \lstick{$\ket{i(0)}$} & \qw \gategroup[wires=13,steps=1,style={dashed,rounded corners,inner xsep=1pt},label style={yshift=0.0cm}]{\text{init.}} & \gate[wires=2]{H_{M}(0)} \gategroup[wires=13,steps=6,style={dashed,rounded corners,inner xsep=1pt},label style={yshift=0.0cm}]{\text{1 Trotter step}} & \gate[wires=5]{H_{I}^{(1)}(0)} & \qw & \gate[wires=2]{H_{E}^{(3)}(0)} & \qw & \gate[wires=5]{H_{I}^{(2)}(0)} & \qw \\
        \lstick{$\ket{o(0)}$} & \gate{X} &  & & \qw & & \qw & & \qw \\
        \lstick{$\ket{i(1)}$} & \qw & \gate[wires=2]{H_{M}(1)} & & \gate[wires=8]{H_{I}^{(2)}(1)} & \gate[wires=3]{H_{E}^{(3)}(1)} & \gate[wires=8]{H_{I}^{(1)}(1)} & & \qw \\
        \lstick{$\ket{o(1)}$} & \gate{X} &  & & & & & & \qw \\
        \lstick{$\ket{l(1)}$} & \gate{X} & \gate{H_{E}^{(1)+(2)}(1)} & & & & & & \qw \\
        \lstick{$\ket{i(2)}$} & \qw & \gate[wires=2]{H_{M}(2)} & \gate[wires=8]{H_{I}^{(1)}(2)} & & \gate[wires=5]{H_{E}^{(3)}(2)} & & \gate[wires=8]{H_{I}^{(2)}(2)} & \qw \\
        \lstick{$\ket{o(2)}$} & \qw & & & & & & & \qw \\
        \lstick{$\ket{l_0(2)}$} & \qw & \gate[wires=2]{H_{E}^{(1)+(2)}(2)} & & & & & & \qw \\
        \lstick{$\ket{l_1(2)}$} & \gate{X} &  & & & & & & \qw \\
        \lstick{$\ket{\text{anc.}}$} & \qw & \qw & & & & & & \qw \\
        \lstick{$\ket{i(3)}$} & \gate{X} & \gate[wires=2]{H_{M}(3)} & \qw & \qw & \qw & & \qw \\
        \lstick{$\ket{o(3)}$} & \qw &  & & \qw & \qw & \qw & & \qw \\
        \lstick{$\ket{l(3)}$} & \gate{X} & \qw & & \qw & \qw & \qw & & \qw 
    \end{quantikz}
    \caption{First-order Trotterized circuit for 4 lattice sites. Each labeled block represents the propagator associated with the indicated Hamiltonian subterm rather than an elementary gate. Blocks spanning multiple register lines indicate the qubits on which the corresponding subterm acts; they do not necessarily imply entangling operations between all of those qubits.}
    \label{fig:1stepcircuit}
\end{figure*}
The Hamiltonian is Trotterized to first order to simulate real-time evolution~\cite{Lloyd:1996aai,Childs:2019hts,Hariprakash:2023tla,Ostmeyer:2026ckx}.
Throughout the hardware simulations, we set $\mu=1$ and $x=1$ in Eqs.~\eqref{eq:mass-term-LSH}--\eqref{eq:interaction-term-LSH} for simplicity.
Each site consists of two fermionic registers, $n_i(r)$ and $n_o(r)$, and one bosonic register, $n_l(r)$. 
For the $N_s=4$ system and boundary conditions considered here, each $n_l(r)$ register requires at most two qubits.
The quantum circuit is pre-optimized based on the boundary conditions. 
For example, we do not need a qubit for $n_l(0)$ because $n_l(0)$ is never greater than 0, allowing the ladder operators associated with that quantum number to also be simplified.

Fig.~\ref{fig:1stepcircuit} shows an overview of the circuit for a single Trotter step applied to an initial state. 
The registers~$\ket{i(r)}, \ket{o(r)}, \ket{l(r)}$ encode the values of~$n_i(r), n_o(r), n_l(r)$, respectively. 
For $r=2$, the bosonic register~$\ket{l(r)}$ is represented as bit strings:
\begin{equation}
    n_l(r=2) = l_0(r=2) + 2\cdot l_1(r=2)\,,
\end{equation}
where $l_0$ and $l_1$ denote the least- and most-significant bits, respectively, so that $\ket{l_1l_0}=\ket{00}$, $\ket{01}$, and $\ket{10}$ encode $n_l(2)=0,1,$ and 2.
Because $\ket{l_1 l_0}=\ket{11}$ lies outside the physically accessible subspace (see Appendix~\ref{app:preoptimization}), the diagonal phase of $H_E^{(1)}+H_E^{(2)}$ at $r=2$ needs only to be correctly produced on $\ket{00}$, $\ket{01}$, and $\ket{10}$.
Moreover, we introduce one ancilla qubit to simplify the ladder operator associated with $n_l(2)$, resulting in a total of 13 qubits in our circuit.
Fig.~\ref{fig: lattice dof}(b) depicts all qubits except for the ancilla.

More generally, we exploit the AGL constraints to choose, for each hopping subterm, an equivalent circuit decomposition that reduces the support of the SVD-diagonalized operator. In particular, whenever possible, the circuitization is chosen so that the two-qubit $n_l(2)$ register does not enter the central diagonal block. We also make use of the freedom in the SVD-control qubit. Details and diagnostic comparisons of these choices are given in Appendix~\ref{app:preoptimization}.

Within each first-order Trotter step, the operators are applied in the left-to-right order shown in Fig.~\ref{fig:1stepcircuit}.
The ordering of terms is designed to minimize the circuit depth and allow ladder operators to cancel out between $H_I^{(1)}(2)$ and $H_I^{(2)}(1)$, as well as between $H_I^{(1)}(1)$ and $H_I^{(2)}(2)$.
The pre-optimized circuit for a single Trotter step is compiled together with the initial-state preparation as a single unit using pytket for execution on the Quantinuum System Model H2.
This joint compilation allows additional gate cancellations involving the preparation circuit. The compiled structure is then used to construct circuits with the desired number of Trotter steps, as detailed in Appendix~\ref{app:preoptimization}. 
The resulting circuit contains 226 two-qubit (\texttt{ZZPhase}) gates per Trotter step.
To maximize hardware resources, we adaptively adjusted the number of shots between 100 and 540, depending on the circuit depth; more shots were allocated to deeper circuits that suffer more from noise.

\subsection{Error mitigation and post-selection}
\label{subsec:QEM}
We adopted post-processing methods to mitigate the effects of noise: (i) a rescaling method assuming a depolarizing noise model and (ii) a post-selection procedure that uses Gauss's law as an error-detection mechanism.
We also considered several other error-mitigation and error-suppression techniques, but did not adopt them either because preliminary tests showed negligible improvement or because their expected experimental overhead was prohibitive for the available hardware resources. 
These alternatives are briefly discussed in Appendix~\ref{app:other-qem}.

The first method we incorporated, depolarizing-model rescaling, has been developed as a general error-mitigation strategy~\cite{Vovrosh:2021ocf,Urbanek:2021oej}.
Related rescaling and self-mitigation methods have also been applied and extended in quantum simulations of lattice gauge theories~\cite{ARahman:2022tkr,Farrell:2022wyt,Ciavarella:2023mfc,Farrell:2023fgd}.
Assuming a depolarizing noise model, the noisy expectation value $\langle O \rangle_{\mathrm{noisy}}$ is related to the ideal expectation value $\langle O \rangle_{\mathrm{ideal}}$ through
\begin{align}
\braket{O}_{\text{noisy}} &=
    (1-p)\braket{O}_{\text{ideal}} +\frac{p}{2^n}\Tr[O]
    \\
    &=(1-p)\braket{O}_{\text{ideal}} + o_{\text{tr}}p\,, \label{eq:depmodel}
\end{align}
where $o_{\text{tr}}\coloneq\Tr[O]/2^n$. 
When the observable~$O$ is expressed as a sum of Pauli operators, only the term proportional to the identity contributes to the trace. Therefore, the~$o_{\text{tr}}$ can be computed straightforwardly. 
By estimating the effective depolarizing parameter~$p$ from calibration data, Eq.~\eqref{eq:depmodel} can be inverted to obtain the error-mitigated expectation value as
\begin{align}\label{eq:QEM}
    \braket{O}_{\text{mit}}=\frac{\braket{O}_{\text{raw}}-o_{\text{tr}}}{1-p}+o_{\text{tr}}\,.
\end{align}
We explored several methods for estimating the depolarizing parameter and selected the most appropriate one, with details of the emulator-based study provided in the Appendix~\ref{app:depnoisestudy}.
The approach we choose introduces an effective depolarizing parameter $p_i$ for each observable $O_i$.
This accounts for the fact that different observables are affected by different levels of accumulated noise because they probe different subsets of qubits and gate sequences.

Gauge constraints have also been explored as a resource for error detection and symmetry-based error mitigation in quantum simulations of lattice gauge theories~\cite{Stryker:2018efp,Klco:2019evd,Raychowdhury:2018osk,Rajput:2021trn,Ballini:2024qmr,Stannigel:2013zka,Lamm:2020jwv,Halimeh:2021vzf,Nguyen:2021hyk,Carena:2024dzu,Modi:2026wte,Cobos:2025krn,Cochran:2024rwe,Sukeno:2026vjm,Mildenberger:2022jqr}. In the LSH formulation used here, the remaining Abelian Gauss's law constraints can be checked directly from the measured computational-basis bit strings. Post-selection therefore consists of keeping or discarding a measured bit string depending on whether it satisfies these constraints.
For each link $r$ connecting sites $r$ and $r+1$, one should have $\hat{N}^{L}(r) = \hat{N}^{R}(r)$.
At the algorithmic gate-decomposition level, the time evolution is designed to exactly conserve the gauge constraints.
In practice, imperfect hardware can introduce detectable violations. 
Since the constraints depend only on the measured bit strings, violations can be detected directly from the measurement outcomes. A common source of such violations is bit-flip errors. Because each quantum number enters two neighboring AGL constraints, an isolated bit-flip error typically causes the violation of two constraint checks and can therefore be detected effectively through post-selection.

On the surface, the constraints do not seem to be a probe for phase errors because the phase on any final-state computational-basis component is entirely separate from the bit string it represents.
Said differently, the constraints are a function of qubit Pauli-$Z$ operators alone, so $Z$ errors commute with the constraints.
However, intermediate changes of basis that take place during the Trotter step do not, in general, commute with qubit $Z$ operators, so they can have the effect of transforming mid-circuit $Z$ errors into $X$ or $Y$ errors.
Thus, AGL post-selection can detect a broader class of circuit errors than computational-basis bit flips alone, including certain mid-circuit phase errors after their propagation through basis-changing operations.

\begin{figure*}
    \centering
    \includegraphics[width=0.8\linewidth]{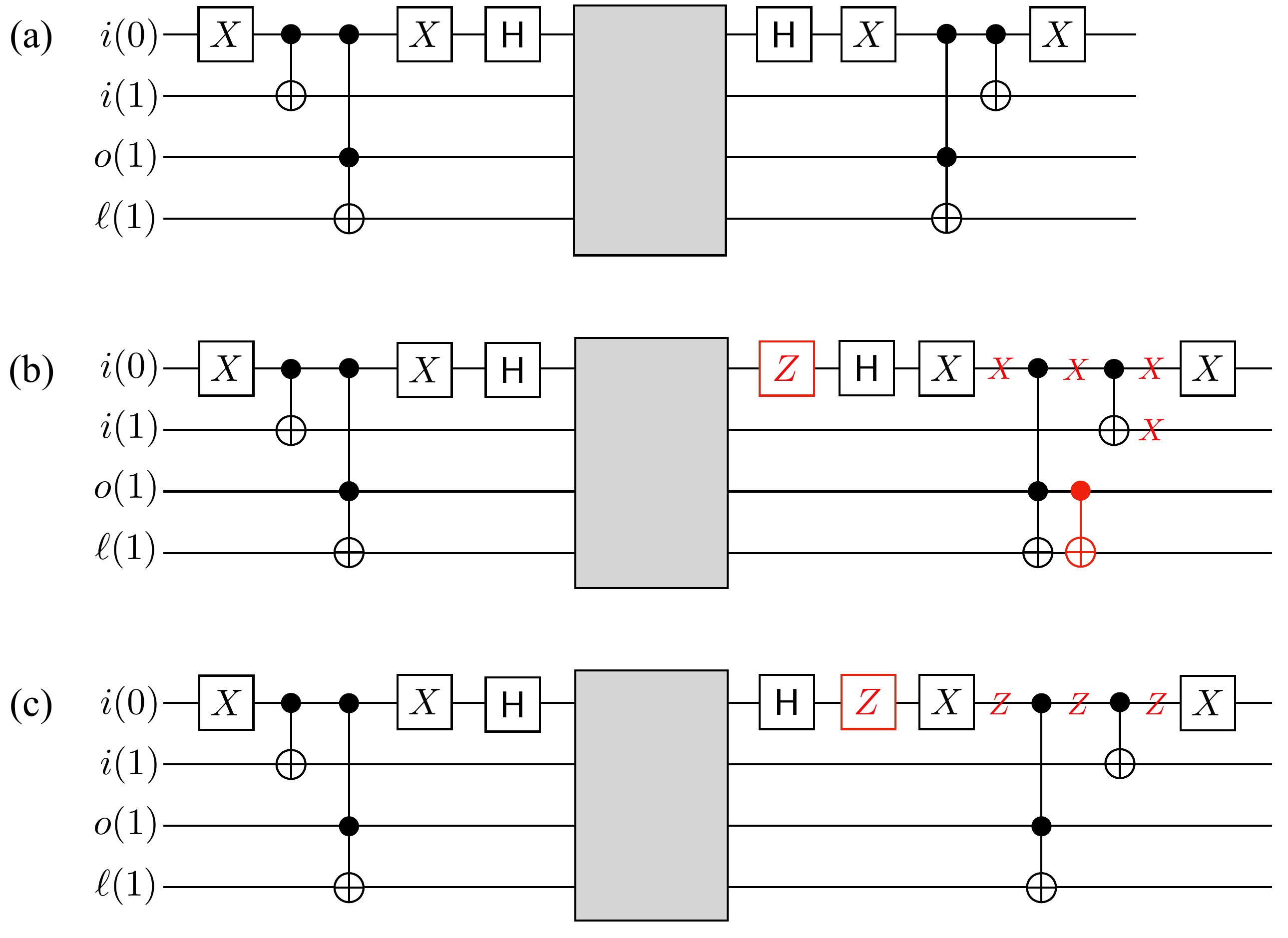}
    \caption{An example of error conversion in the hopping subterm $H^{(j=1)}_I(r=0)$. Registers $i(r),o(r),\ell(r)$ are assigned for $n_i(r), n_{o}(r), n_{\ell}(r)$, respectively. The gray box represents the diagonal block, implemented using $R_Z$ and $CX$ gates. (a) The original circuit without the error insertion. (b) A $Z$-error between $\mathsf{H}$ gates is converted into the $X$ error, which is further propagated and spreads out. (c) A $Z$-error outside two $\mathsf{H}$ gates propagates and can change some angles, but is not converted to $X$-errors, at least in this subcircuit.}
    \label{fig:circ-error-conversion}
\end{figure*}

Indeed, when implementing the Trotter evolution for hopping subterms, the diagonalized forms are sandwiched by two Hadamard gates in $\mathcal{U}_{\text{SVD}}$ and $\mathcal{U}_{\text{SVD}}^\dag$, as shown in Eqs.~\eqref{eq: svd-unitary 1}-\eqref{eq: svd-unitary 2}. 
Because the diagonal block is implemented using $R_Z$ and $CX$ gates, $Z$ errors remain $Z$-type while propagating through the block, although they may spread to additional qubits. A subsequent Hadamard gate can then convert such an error into an $X$ error detectable by the AGL constraints.
We give an explicit example in Fig.~\ref{fig:circ-error-conversion}. 
The circuit in Fig.~\ref{fig:circ-error-conversion}(a) shows the time evolution for the hopping subterm $H^{(j=1)}_I(r=0)$.
The registers $i(r),o(r),\ell(r)$ correspond to $n_i(r), n_{o}(r), n_{\ell}(r)$, respectively. The gray box represents the diagonal part, which is composed of $R_Z$ and $CX$ gates. Note that the gray box has a much longer depth than the other parts in this circuit.
As shown in Fig.~\ref{fig:circ-error-conversion}(c), one can see that $Z$ errors outside the two $\mathsf{H}$ gates cannot be converted into $X$ errors and are therefore invisible via the AGL, although they can change some phases and affect final measurements.
On the other hand, the $Z$ error inserted between two $\mathsf{H}$ gates as in Fig.~\ref{fig:circ-error-conversion}(b) becomes an $X$ error and thus can be detected by the AGL measurement. Moreover, the $X$ error will further propagate and spread out via $CX$ and Toffoli gates. This can potentially increase the sensitivity to this type of error, albeit at the cost of increasing the discard rate. 

\section{\label{sec: results}Results}
In this section, we report the results obtained from time evolution on the hardware and compare them against the idealized zero-noise evolution. The benchmark values are regarded as those of noiseless Trotter evolution, as opposed to continuous-time evolution.
We use a fixed Trotter timestep of $\Delta t=1.25$, which allows the experimentally accessible six Trotter steps to span a time range over which the noiseless Trotterized electric-energy densities exhibit clear nontrivial oscillatory behavior.
Trotterization effects relative to the continuous-time evolution, as well as exploratory approaches for reducing them, are discussed separately in Appendix~\ref{app:trotterstudy}.
We execute circuits with up to six Trotter steps, corresponding to a maximum evolution time of $t=7.5$, within the available hardware-resource and circuit-depth constraints.
Following an explanation of our error mitigation procedures, we analyze the scaling of the effective noise parameter and post-selection discard rate with circuit depth. 
Finally, the mitigated results are presented and compared to the noiseless baseline.

\subsection{Calibration for the depolarizing error model}
\begin{figure}
    \centering
    \includegraphics[width=0.95\linewidth]{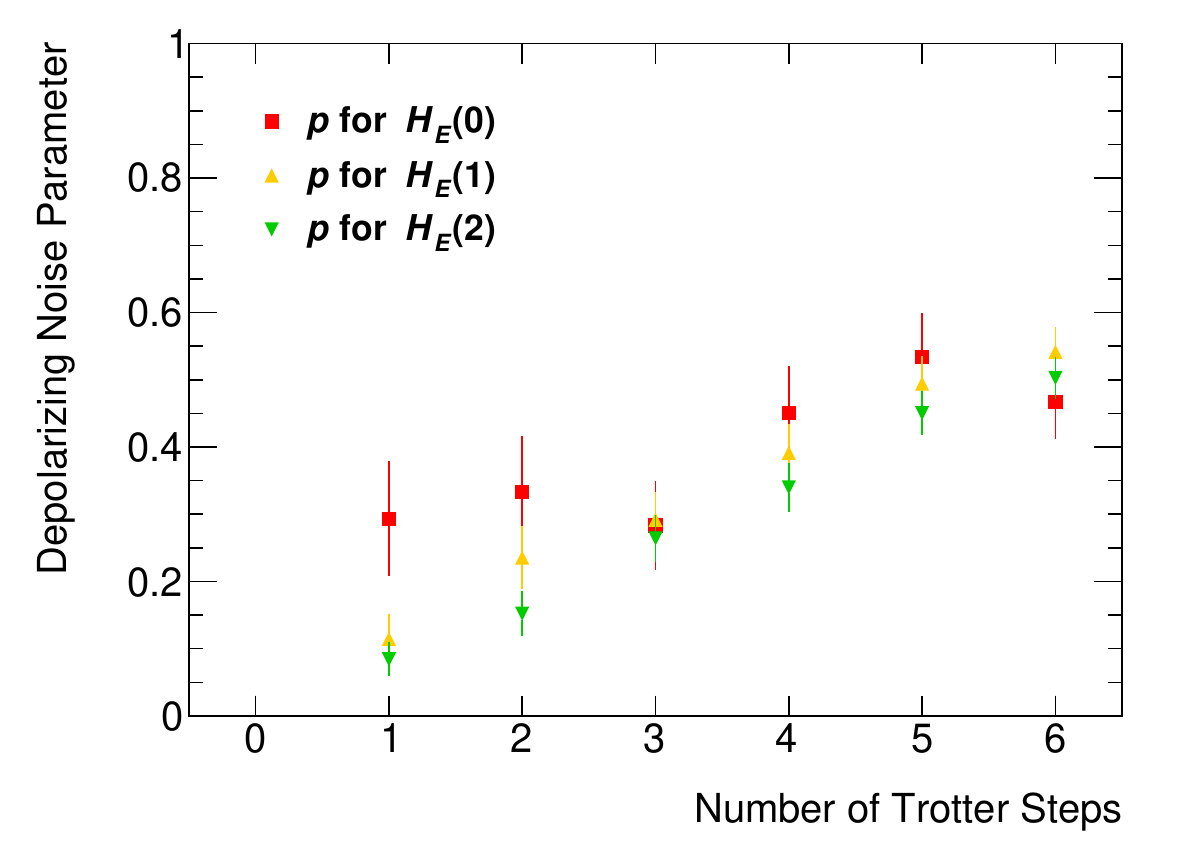}
    \caption{Estimated effective depolarizing-noise parameters obtained using the test circuits. The red squares, yellow triangles, and green inverted triangles correspond to the observables on links 0, 1, and 2, respectively.}
    \label{fig:depnoise_p}
\end{figure}

To apply depolarizing error mitigation, the effective error parameter $p_{i}$ must be estimated independently of the observables of interest. 
We use quantum circuits in which all $n_i(r)$, $n_o(r)$, and $n_l(r)$ in the initial state are set to zero as test circuits in this study.
In this test circuit,
the noiseless expectation value in Eq.~\eqref{eq:depmodel} is $\braket{O_i{}}_{\text{ideal}}=0$. Therefore, the noise parameter~$p_{i}$ can be estimated as $p_{i}=\braket{O_{i}}_\mathrm{raw}^\mathrm{test}/o_{\mathrm{tr},i}$, where $\braket{O}_\mathrm{raw}^\mathrm{test}$ is measured expectation value obtained from the test circuit.
Obtained noise parameters for each observable are shown in Fig.~\ref{fig:depnoise_p}.
The horizontal axis represents the number of Trotter steps, and therefore the circuit depth also increases along this axis.
The overall increase in the noise parameter with circuit depth suggests that our depolarizing noise model is not significantly inconsistent with the observed behavior.
Depolarizing error-mitigated values for each observable are calculated according to Eq.~\eqref{eq:QEM} with the estimated $p_{i}$ in each Trotter step.

\subsection{Discard rate with post-selection}
\begin{figure}
    \centering
    \includegraphics[width=1.0\linewidth]{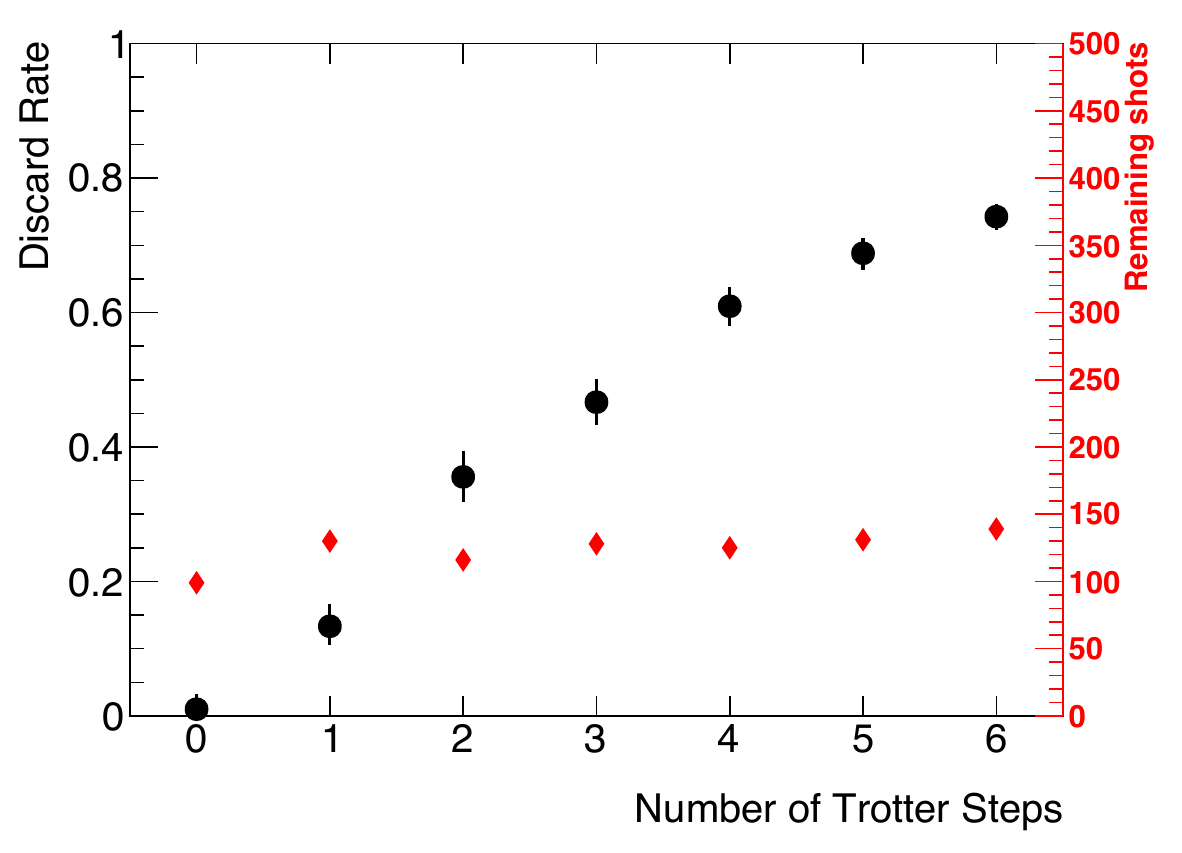}
    \caption{Discard rate and number of shots remaining after post-selection as functions of the number of Trotter steps. The black left vertical axis corresponds to the discard rate shown by black circles, while the red right vertical axis corresponds to the number of remaining shots shown by red diamonds.}
    \label{fig:discard-rate}
\end{figure}

Instead of employing the depolarizing error mitigation, one can use the post-selection methods. 
To be clear, the depolarizing error mitigation and post-selection we used are \emph{not} implemented simultaneously.
As a post-selection, we keep results that satisfy the Abelian Gauss's law:~$\hat{N}^{L}(r) = \hat{N}^{R}(r)$ for $0 \le r \le 2$. 
We also require the ancilla qubit to return to $\ket{0}$, the left boundary condition $\hat{N}^{R}(-1)=0$, and the right boundary condition $\hat{N}^{L}(3)=1$ to be satisfied.
We furthermore impose the matter-superselection-sector condition of Eq.~\eqref{eq: matter superselection sector}.
The impact of the AGL requirement is dominant compared with the ancilla and matter-superselection-sector conditions.
If the results do not satisfy the above conditions, we can be sure that an error occurred during the computation.
By discarding such outcomes, we can mitigate errors without assuming an explicit noise model. However, this procedure detects only errors that violate at least one of the imposed physical constraints; errors that preserve all constraints remain in the post-selected data.
The discard rate, defined as the ratio of discarded shots to the total number of shots, and the number of shots remaining after post-selection are shown in Fig.~\ref{fig:discard-rate} as functions of the number of Trotter steps.
The number of shots was adjusted between 100 and 540 depending on the circuit depth, so that the number of accepted shots after post-selection remained approximately in the range of 100--150 while minimizing the required hardware resources.
The discard rate increases with circuit depth, reaching 0.74 at six Trotter steps.

\subsection{Error-mitigated results}
\begin{figure}[p]
    \centering
    \includegraphics[width=1.0\linewidth]{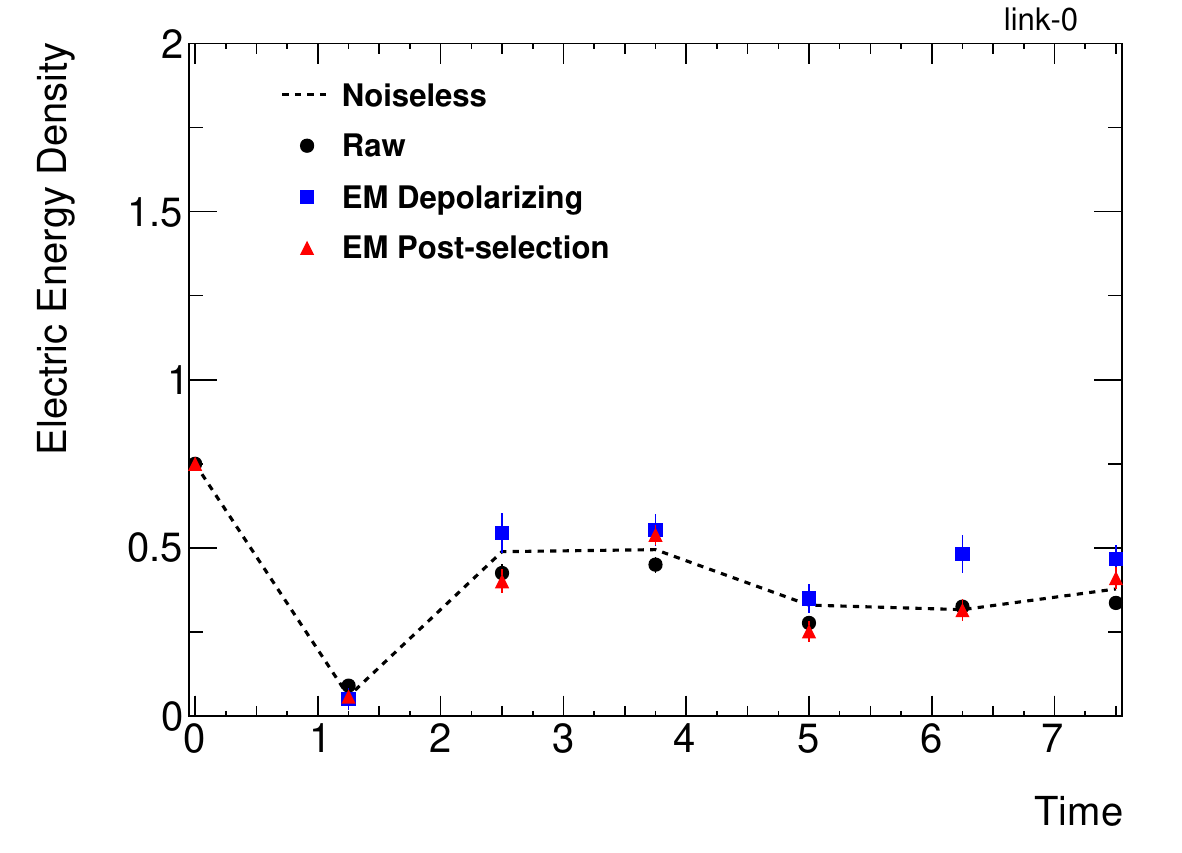} \label{fig:obs-h2-site0} \\
    \includegraphics[width=1.0\linewidth]{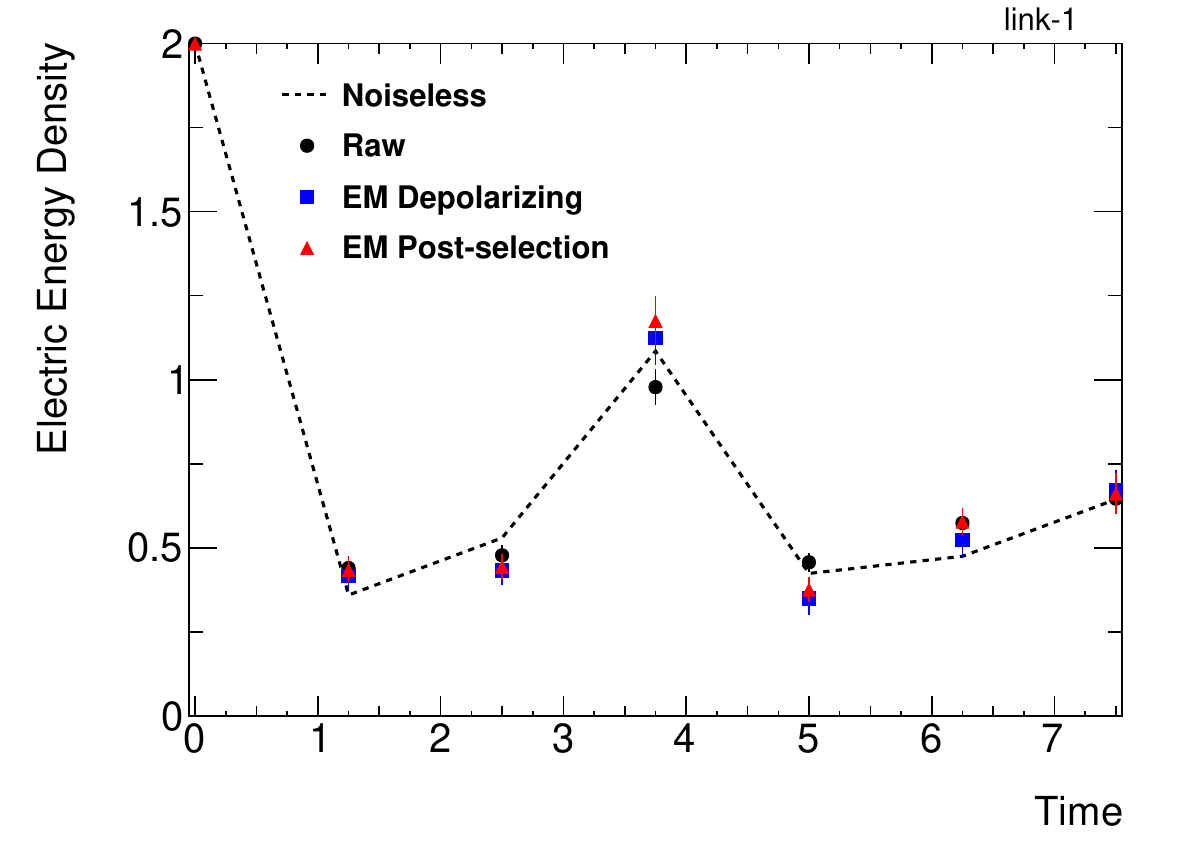} \label{fig:obs-h2-site1} \\
    \includegraphics[width=1.0\linewidth]{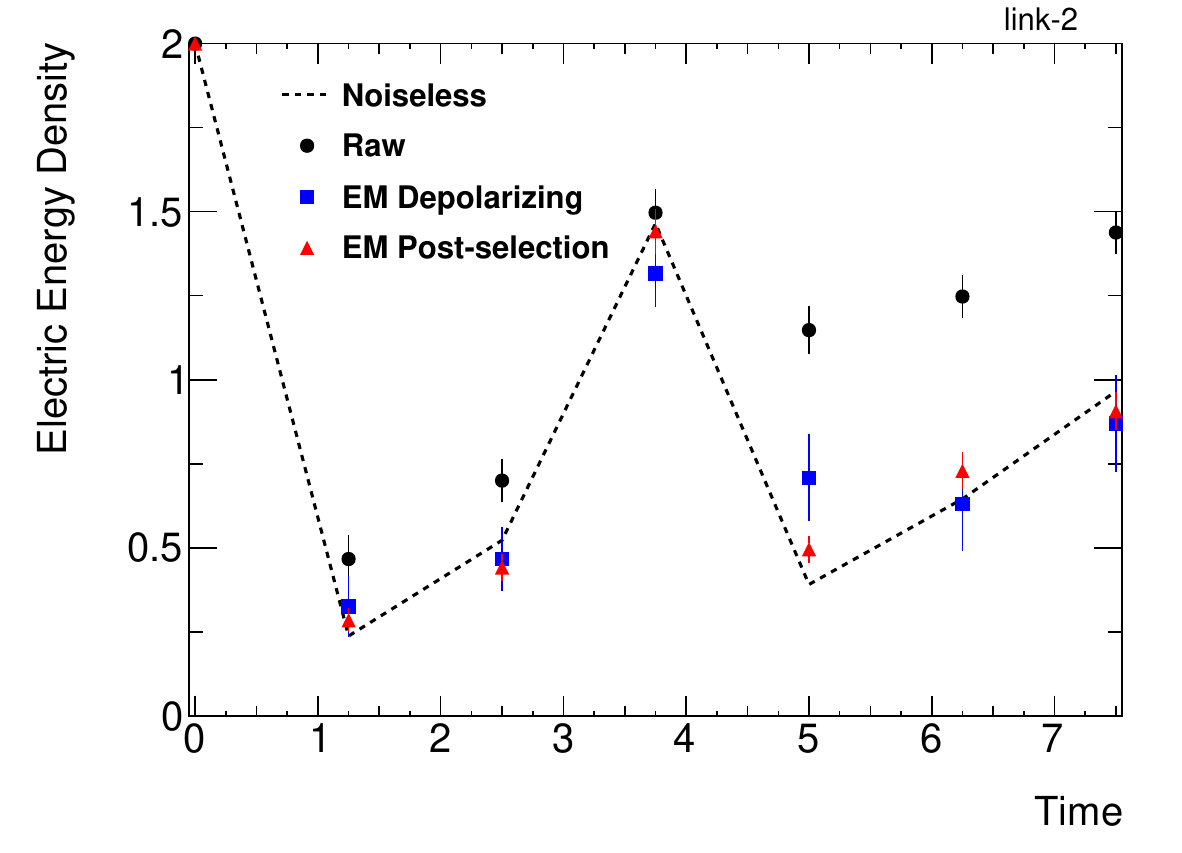} \label{fig:obs-h2-site2}
    \caption{Electric energy density at each lattice link. The horizontal axis represents dimensionless time, and the vertical axis represents electric energy density. The timestep is fixed at 1.25, and the circuit depth (i.e., the number of Trotter steps) scales linearly with time. The dotted line represents the noiseless simulation result, and the markers are the results on the Quantinuum H2-2 device. The black circles are observed raw values, the blue squares are depolarizing error-mitigated values, and the red triangles are the error-mitigated values using post-selection.}
    \label{fig:elec-each-site-QEM_hard}
\end{figure}

Fig.~\ref{fig:elec-each-site-QEM_hard} shows the electric-energy density for the hardware as a function of total evolution time, where the dynamics are decomposed into Trotter steps with a fixed timestep size $\Delta t = 1.25$.
In this and subsequent figures and tables, ``EM'' denotes error-mitigated results; ``EM Depolarizing'' and ``EM Post-selection'' refer to the results obtained with depolarizing-noise rescaling and post-selection, respectively.
The electric-energy density is defined in Eq.~\eqref{eq:obs}, and is therefore evaluated directly from the measured $(n_i(r), n_o(r), n_l(r))$ bitstrings. 
Observed values are compared with the noiseless Trotterized-circuit baselines.
Error bars represent statistical uncertainties. For the raw and post-selected results, the uncertainty of each observable is estimated as $\sigma/\sqrt{N_{shots}}$, where $\sigma$ is the standard deviation of the shot-by-shot observable values and $N_{shots}$ is the number of shots used in the corresponding estimate. Although post-selection reduces $N_{shots}$, it also reduces $\sigma$ in the present data by removing outcomes that violate the imposed physical constraints. Consequently, the statistical uncertainties of the post-selected results remain comparable to those of the raw data despite the smaller number of retained shots. For the depolarizing-rescaled results, the statistical uncertainty in the independently estimated effective noise parameter is additionally propagated.

For links 0 and 1, even the raw results are already close to the noiseless Trotterized baselines over the full range of circuit depths. The mitigation corrections are therefore relatively modest. For link 0, however, the depolarizing-model-rescaled values are systematically shifted upward, suggesting a slight overestimation of the effective depolarizing parameter. For link 1, the corrected values remain broadly consistent with the baseline within the statistical uncertainties.

Link 2 is the most challenging observable. The electric energy density on link 2 depends on more qubits than any other link, involving the two-qubit $n_l(2)$ register that explores three distinct states under idealized evolution. Consequently, a bit-flip error in the $n_l(2)$ register can produce a relatively large change in the reconstructed electric-energy value. In addition, the expectation values for a uniform random distribution differ among the three links and are approximately 0.19, 0.63, and 2.0 for links 0, 1, and 2, respectively. Therefore, link 2 is affected not only by a larger accumulated noise contribution but also by a larger separation between the noiseless expectation value and the random-distribution limit.

Accordingly, the raw data for link 2 show clear deviations from the noiseless Trotterized baseline, particularly for the deeper circuits. Nevertheless, even for the deepest circuit considered here, namely the six-Trotter-step circuit, the raw values remain far from the uniform-random limit of 2.0. Likewise, the effective depolarizing parameters and the post-selection discard rates remain below their corresponding random-distribution limits of 1.0 and 0.99, respectively. 
The present experiment was limited to six Trotter steps by the available hardware-execution budget. At this maximum depth, the measured observables remain well separated from their fully mixed-state limits, indicating that the hardware results still retain meaningful information and suggesting that useful results may remain obtainable at even longer evolution times.

Both error-mitigation methods, namely depolarizing-noise rescaling and post-selection, produce measured values that are close to the noiseless limit. 
The electric-energy density on link 2, for which the raw-data deviations were largest, experiences the most improvement from error mitigation.

In the remainder of this section, we offer some figures of merit to quantify the closeness of the data to the noiseless Trotterized-circuit baseline.
Relevant quantities to consider are the error-mitigated expectation values $\braket{H_{E}(r, t)}_{\rm data}$, the noiseless-limit expectation values $\braket{H_{E}(r, t)}_{\rm ref}$, and the uniform-random mixed-state expectation values $\braket{H_{E}(r, t)}_{\rm mixed}$.
We compute, in particular, the differences
\begin{align}
    \Delta H_E(r, t) &= \braket{H_{E}(r, t)}_{\rm data} - \braket{H_{E}(r, t)}_{\rm ref}.
\end{align}
As a point of reference, the deviations $|\Delta H_E(r, t)|$ can be compared to $ | \braket{H_{E}(r, t)}_{\rm mixed} - \braket{H_{E}(r, t)}_{\rm ref} |$, which represents a type of worst-case-scenario deviation.

\begin{figure*}
    \centering
    \includegraphics[width=0.45\linewidth]{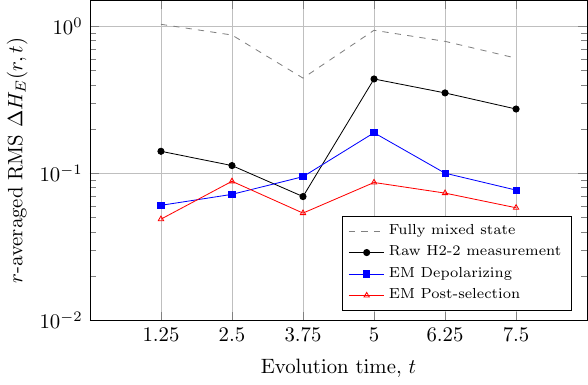}
    \hspace{.05\linewidth}
    \includegraphics[width=0.45\linewidth]{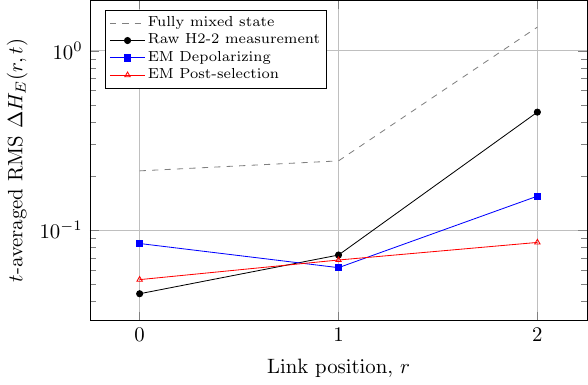}
    \caption{Averaged deviations in $H_E(r,t)$ relative to a noiseless simulation. The left (right) panel shows the root-mean-square deviations obtained by averaging over $r$ at fixed values of $t$ ($t$ at fixed values of $r$).}
    \label{fig: electric deviations}
\end{figure*}

The deviations can be aggregated into root-mean-square (RMS) deviations in a few ways.
One can fix $t$ and compute the RMS deviation by averaging over $r$, or fix $r$ and average over the sampled values of $t$. Hence we define
\begin{align}
    \rm{RMS}_r(t) &= \sqrt{\frac{1}{(N_{s}-1)}\sum^{N_s-2}_{r=0} [\Delta {H_E(r, t)}]^{2}}, \label{eq:rms at t} \\
    \rm{RMS}_t(r) &= \sqrt{\frac{1}{N_t}\sum_{t} [\Delta {H_E(r, t)}]^{2}}, \label{eq:rms at r}
\end{align}
with $N_t$ being the number of sampled values of $t$.
These two perspectives are plotted in Fig.~\ref{fig: electric deviations}.
For our point of reference, the same quantities are plotted for the completely mixed state (the dashed lines).
These plots support the conclusion that error mitigation tends to substantially improve the accuracy, with depolarizing-noise and post-selection methods giving comparable improvements but post-selection tending to be slightly better. One observation we make is that in cases where the raw measurement is already close to the reference value, the depolarizing-noise rescaling method risks increasing the error: this is seen at the $t=3.75$ time slice or for the link $r=0$. The same risk exists for the post-selection method, as is observed for link $r=0$; however, the effect is relatively small.

As a further consolidation, one can use the central values obtained for all $(r,t)$ pairs in a data series and compute the RMS deviation for the series as a whole:
\begin{align}
    \rm{RMS}_{\rm full} &= \sqrt{\frac{1}{(N_s - 1) N_t}\sum^{N_s-2}_{r=0} \sum_{t} [\Delta {H_E(r, t)}]^{2}} . \label{eq:rms total} 
\end{align}
These numbers are reported in Table \ref{tab: consolidated deviations}.
The uncertainties are statistical and are obtained by propagating the finite-shot uncertainties of the individual observables.
For the depolarizing-rescaled results, the statistical uncertainty in the independently estimated effective parameter $p_i$ is also propagated.
This metric continues to affirm the meaningful information extracted from raw data alone and the significant improvements possible with error mitigation, although it misses some of the subtleties.
\begin{table}[]
    \centering
    \begin{tabular}{l c}
        & $r$- and $t$-averaged RMS deviation \\
        \hline
        Fully mixed state & 0.806 \\
        Raw H2-2 data & 0.267  $\pm\ 0.016$ \\
        EM Depolarizing & 0.108  $\pm\ 0.025$ \\
        EM Post-selection & 0.067  $\pm\ 0.011$
    \end{tabular}
    \caption{Root-mean-square deviations $\Delta H_E(r,t)$ from the noiseless Trotterized-circuit results, averaged over all lattice links and sampled evolution times.}
    \label{tab: consolidated deviations}
\end{table}

According to the RMS metric in Eq.~\eqref{eq:rms total}, deviations of the electric-energy density in the raw data tend to be about $30\%$ of their size in the worst-case limit.
With error mitigation methods, those same deviations are reduced to about $10\%$.
Alternatively, we can say that the overall RMS deviation of the raw data compared to the worst-case scenario is improved by about 60\% with depolarizing-noise rescaling (0.108 compared to 0.267) and about 75\% with post-selection (0.067 compared to 0.267).

\section{\label{sec: discussion}Discussion}
We have presented the results of a quantum simulation for the (1+1)-dimensional SU(2) lattice gauge theory with fermionic matter.
Using a fixed Trotter-step time interval for up to six time steps, we simulated the time evolution of the electric-energy-density profile throughout the four-staggered-site lattice.
The raw data tended to follow the noiseless Trotterized-circuit baseline most closely in the regions that were expected to be less impacted by noise, i.e., the first two links of the three-link lattice.
After error mitigation, the data were in approximate agreement with what one would have expected in the limit of zero noise.

Overall, we found post-selection to be the more favorable mitigation strategy for the present study. Although its fully consolidated RMS deviation is only moderately smaller than that of depolarizing rescaling within the present statistical uncertainties, post-selection also yields smaller statistical uncertainties, requires no auxiliary noise-calibration circuits or explicit noise model, and avoids the systematic shift observed for the depolarizing-rescaled link-0 results. Its main drawback is the increasing shot-discard rate with circuit depth.
We emphasize that this is a qualitative assessment: with the shot budget used here, the consolidated deviations of the two mitigated data sets in Table~\ref{tab: consolidated deviations} differ by roughly $1.5$ combined standard deviations.

In this theory, and in the LSH formulation, the relevant local dynamical constraints are the Abelian Gauss's laws, which require flux conservation along any given link.
In contrast to the other method, the post-selection strategy applies to all observables simultaneously.
Furthermore, we have shed light on how post-selection on the Abelian Gauss's law can be sensitive to multiple types of qubit errors.
The observed performance is consistent with the expectation that sensitivity to multiple error types ought to enhance the effectiveness of the post-selection.
This is illustrated by the fact that the data obtained from post-selection closely follow the ideal-simulator results even up to six Trotter steps.

Due to the stringent nature of the post-selection, the discard ratio is an important consideration.
To that end, before running on hardware, we used the emulator to estimate the discard rates for each number of Trotter steps. Hence, we could allocate real hardware shots such that $O(100)$ shots should pass post-selection for all considered evolution times.
For our deepest circuits of six Trotter steps, 26\% of the shots passed post-selection, hence we could expect to accumulate meaningful statistics within our simulation budget.
However, under approximately local independent errors, the probability that all constraints are simultaneously satisfied can decrease exponentially with the number of constraints, leading to a rapidly increasing sampling overhead.
For larger-scale simulations, one would have to weigh the drawback of high discard rates against the advantage offered by seemingly effective error mitigation and quality of data from the shots that do pass the test.

As a scalable error-mitigation strategy, we applied the depolarizing noise model for each tracked observable.
For raw data that was originally far off from the expected values, the depolarizing-noise model tended to improve the signal.
However, for data points whose raw values were already close to the idealized-simulator value, this error mitigation strategy sometimes led to data that were farther off from the idealized value.
Thus, this strategy may be used to overall improve the data, but the statistical uncertainties can be substantially amplified.

We should point out that our EM methods are known not to be exhaustive, so some disagreement between the final data and the noiseless-limit data is to be expected. The fact that the EM data fall close to the noiseless limit using only the simplistic methods we have presented is encouraging and suggests that complete statistical agreement could be within reach of more advanced EM methods.

\section{\label{sec: conclusion}Conclusion}
In this work, we demonstrated a real-time quantum simulation of a (1+1)-dimensional SU(2) lattice gauge theory with dynamical fermions on Quantinuum System Model H2 trapped-ion processor using the loop-string-hadron formulation. 
By retaining the local bosonic gauge degrees of freedom and encoding $n_{l}\in\{0,1,2\}$ in a two-qubit register, we accessed the $j=1$ electric-flux sector beyond the commonly used $j_{\mathrm{max}}= 1/2$ cutoff. 
This constitutes a minimal hardware demonstration of a theory synthesizing several key features: a non-Abelian gauge group, enforced gauge invariance, qubit encodings for both matter fields and local gauge degrees of freedom, a multiqubit register for the bosons, and the simulation of dynamics beyond the harshest nontrivial electric-basis truncation.
All of these features are relevant to the escalation of quantum simulations from the simplified models that can be simulated today to what will one day be large-scale simulations of QCD.

Our simulation involved first-order Trotterized time-evolution circuits for up to six steps, by which we could study the real-time evolution of the electric-energy density on all three links. 
Both observable-specific, depolarizing-model rescaling and observable-agnostic, AGL-based post-selection improved the agreement of our data with the corresponding noiseless Trotterized-circuit baselines, with the clearest improvement observed for the most noise-sensitive link-2 observable. 
The results show that fermionic fields and multiqubit gauge-field registers can be implemented simultaneously and meaningfully probed on current quantum hardware. 
Together with the locality of the LSH Hamiltonian, this demonstration provides a foundation for simulations of larger systems, higher electric-field truncations, and eventually higher-dimensional non-Abelian lattice gauge theories~\cite{Zohar:2021nyc,Kadam:2022ipf,Burbano:2024uvn,Ciavarella:2024fzw}.

\section*{Acknowledgments}
We thank Michael Foss-Feig and Yuta Kikuchi for their feedback on the manuscript.
We also thank Quantinuum for providing access to the Quantinuum System Model H2 used in this work.
This work was supported by the U.S.–Japan Science and Technology Cooperation Program in High Energy Physics, through the U.S. Department of Energy and the High Energy Accelerator Research Organization (KEK) in Japan. Support was also provided by the U.S. Department of Energy (DOE), Office of Science under contract DE-AC02-05CH11231, partially through Quantum Information Science Enabled Discovery (QuantISED) for High Energy Physics (KA2401032).
This work was also supported by JST PRESTO Grant Number JPMJPR25F5 and JST Grant Number JPMJPF2221.

\section*{Data Availability}
The data that support the findings of this study are available from the corresponding author upon reasonable request.

\appendix

\section{Boundary-condition-aware optimization of the hopping circuits} \label{app:preoptimization}
\subsection{AGL-based equivalence of diagonal forms}
The hopping interaction is the most circuit-intensive part of the Hamiltonian considered in this work. 
We implement its propagators using the singular-value-decomposition (SVD) construction of Ref.~\cite{Davoudi:2022xmb}. 
Here we describe additional simplifications that become available after restricting the dynamics to the Abelian-Gauss-law (AGL)-satisfying sector associated with the boundary conditions of our four-site system.

The generalized labels used in the SVD construction are related to the local LSH quantum numbers by
\begin{align}
\begin{array}{c|cc}
 & H^{(1)}_{I}(r) & H^{(2)}_{I}(r) \\
\hline
x' & n_i(r)   & n_o(r+1) \\
x  & n_o(r)   & n_i(r+1) \\
y' & n_i(r+1) & n_o(r)   \\
y  & n_o(r+1) & n_i(r)   \\
p  & n_l(r)   & n_l(r+1) \\
q  & n_l(r+1) & n_l(r)
\end{array}\label{eq:lookup}
\end{align}
With this notation, the AGL condition $N_L(r)=N_R(r)$ can be written in the common form
\begin{align}
p+n_x(1-n_{x'}) = q+n_{y'}(1-n_y). \label{eq:agl_xy}
\end{align}
Each hopping subterm is Hermitian and may be written as
\begin{align}
H_I^{(j)}=A+A^\dagger,
\end{align}
where one directed contribution $A$ contains the fermionic operator $\sigma^+_{y'}\sigma^-_{x'}$.
Following the convention of Ref.~\cite{Davoudi:2022xmb}
\begin{align}
\sigma^+=\ket{0}\bra{1}, \qquad 
\sigma^-=\ket{1}\bra{0},
\end{align}
so $A$ is only nonzero when applied to kets with
\begin{align}
n_{x'}=0,\qquad n_{y'}=1. \label{eq:fermionictransition}
\end{align}
The corresponding fermionic transition is therefore
\begin{align}
\ket{0,1}_{x'y'} \longrightarrow \ket{1,0}_{x'y'}.
\end{align}
For such a nonzero matrix element, substituting Eq.~\eqref{eq:fermionictransition} into the AGL constraint~\eqref{eq:agl_xy} gives
\begin{align}
p+n_x = q+1-n_y,
\end{align}
or equivalently,
\begin{align}
p-(1-n_x)=q-n_y. \label{eq:diagAGL}
\end{align}
The conditional bosonic lowering operators
\begin{align}
(\lambda_q^-)^{n_y}(\lambda_p^-)^{1-n_x}
\end{align}
act before the diagonal factor is evaluated. 
Eq.~\eqref{eq:diagAGL} therefore states that the bosonic occupations entering the $p$- and $q$-based diagonal forms are equal after the corresponding conditional lowering operations.
Defining
\begin{align}
d(\nu;n_x,n_y) = \sqrt{\frac{\nu+1+n_x}{\nu+1+n_y}}
\end{align}
as the corresponding diagonal matrix element of the $D$ operator for bosonic occupation $\nu$, it follows that
\begin{align}
d\!\left(p-(1-n_x);n_x,n_y\right) = d\!\left(q-n_y;n_x,n_y\right).
\end{align}
The reverse hopping process is generated by $A^\dagger$, so the corresponding equality of the reverse matrix elements follows by Hermitian conjugation.

This equivalence concerns the diagonal factors appearing in nonzero hopping matrix elements and should not be interpreted as a pointwise identity between $p$ and $q$ throughout the full AGL-satisfying Hilbert space. 
The argument above establishes the equality of the corresponding diagonal matrix elements of the $D$ factor for nonzero hopping transitions.
The cutoff conditions are constrained by the same AGL relations, as discussed in Ref.~\cite{Davoudi:2022xmb}. In our implementation, the complete $p$- and $q$-based diagonal operators are constructed separately within the AGL- and boundary-condition-restricted basis. Their equivalence on the relevant physical subspace is verified numerically in Sec.~\ref{sec:physicalsubspace}.

\subsection{Two SVD orientations}
The choice of diagonal form is distinct from the choice of SVD-control qubit. Defining
\begin{align}
{\cal T} = (\lambda_q^-)^{n_y} (\lambda_p^-)^{1-n_x},\
\end{align}
the SVD construction with $y'$ as the control qubit presented in Sec.~\ref{sec: background} is
\begin{align}
U_{\rm SVD}^{(y')} = \mathsf{H}_{y'} \left[ \ket{0}\bra{0}_{y'} + \ket{1}\bra{1}_{y'} X_{x'}{\cal T} \right],
\end{align}
which leads to the fermionic diagonal structure
\begin{align}
\ket{1}\bra{1}_{x'}Z_{y'}Z_x.
\end{align}
An equivalent construction can instead use $x'$ as the SVD-control qubit,
\begin{align}
U_{\rm SVD}^{(x')} = \mathsf{H}_{x'} \left[ \ket{0}\bra{0}_{x'} X_{y'}{\cal T} + \ket{1}\bra{1}_{x'} \right],
\end{align}
yielding the fermionic diagonal structure
\begin{align}
\ket{0}\bra{0}_{y'}Z_{x'}Z_x.
\end{align}
To make these two freedoms explicit, let $\mathcal D_{\beta}$ denote the complete constrained diagonal factor, including the diagonal function and cutoff conditions, constructed using bosonic representation $\beta\in\{p,q\}$. The two SVD orientations can then be written schematically as
\begin{align}
U_{\rm SVD}^{(y')} H_I^{(j)} U_{\rm SVD}^{(y')\dagger} &= x\ket{1}\bra{1}_{x'}Z_{y'}Z_x\,{\cal D}_{\beta}, \\
U_{\rm SVD}^{(x')} H_I^{(j)} U_{\rm SVD}^{(x')\dagger} &= x\ket{0}\bra{0}_{y'}Z_{x'}Z_x\,{\cal D}_{\beta}, \qquad \beta\in\{p,q\}.
\end{align}
Thus, the SVD orientation determines the fermionic basis transformation and projector-$Z$ structure, while the choice of $\beta$ determines which AGL-related bosonic register is retained in the constrained diagonal operator.

\subsection{Representation choices and circuit cost} \label{sec:physicalsubspace}
We numerically checked the two SVD orientations combined with the two bosonic representations for the representative subterms $H_I^{(1)}(1)$ and $H_I^{(2)}(1)$. 
For each representative subterm, the full output state was compared for all local computational-basis configurations admitted by the AGL and the relevant boundary restriction. The four implementations agreed to numerical precision.

For this diagnostic comparison only, the diagonal operators were synthesized using the same Gray-code procedure in all four cases and the resulting isolated hopping subterms were compiled to the H2-2 native gate set using pytket. This common synthesis was used to compare the representations on an equal footing and was not used in the hardware experiment.

\begin{table}[t]
    \centering
\begin{tabular}{ccccc}
Subterm	
& \begin{tabular}{c} SVD \\control \end{tabular} 
& \begin{tabular}{c} Bosonic \\rep. \end{tabular} 
& \begin{tabular}{c} Diagonal \\support\end{tabular} & ZZPhase \\ \hline
$H_I^{(1)}(1)$ & $y'$ &	$p$ & 5 & 56 \\
 & $x'$ & $p$ & 5 & 56 \\
 & $y'$ & $q$ & 6 & 94 \\
 & $x'$ & $q$ & 6 & 91 \\
$H_I^{(2)}(1)$ & $y'$ & $q$ & 5	& 55 \\
 & $x'$ & $q$ & 5 & 58 \\
 & $y'$ & $p$ & 6 & 89 \\
 & $x'$ & $p$ & 6 & 90 
\end{tabular}
\caption{Diagnostic comparison of alternative SVD orientations and bosonic representations for two representative hopping subterms. The same Gray-code synthesis was used for all diagonal blocks before compilation to the H2-2 native gate set. These circuits were used only for this comparison and were not the circuits executed on hardware.}
\label{tab:graycomp}
\end{table}

The dominant cost difference originates from the bosonic representation. For $H_I^{(1)}(1)$, the $p$-representation retains $n_l(1)$ rather than the two-qubit $n_l(2)$ register, whereas for $H_I^{(2)}(1)$ the $q$-representation gives the corresponding reduction. In both cases, excluding $n_l(2)$ reduces the diagonal support from six to five physical qubits and produces a substantially larger reduction in the compiled two-qubit-gate count than changing the SVD orientation alone.

For the boundary conditions considered in this work,
\begin{align}
n_l(0)&=0,\\
n_l(1),n_l(3)&\in\{0,1\},\\
\mathrm{and} \ n_l(2)&\in\{0,1,2\}.
\end{align}
Because $n_l(2)\le 2$, the computational-basis state $\ket{11}$, which would encode $n_l(2)=3$, lies outside the physically accessible subspace. We therefore require the synthesized diagonal operator to reproduce the target matrix elements only on the physically allowed states $\ket{00}$, $\ket{01}$, and $\ket{10}$. The unconstrained action on $\ket{11}$ provides additional freedom in the diagonal synthesis and can be exploited to reduce the two-qubit-gate count.

Thus $n_l(2)$ is the only bosonic degree of freedom requiring a two-qubit data register.
Guided by these boundary-condition-specific register sizes, the bosonic representation used for each hopping subterm was chosen so that the two-qubit $n_l(2)$ register does not enter the central diagonal block whenever possible. The choices used in the hardware circuit are summarized in Table~\ref{tab:choice}.

\begin{table}[t]
    \centering
\begin{tabular}{cccc}
Subterm         & SVD control & Bosonic rep. & Retained register \\ \hline
$H_I^{(1)}(0)$ & $x'$      & $q$        & $n_l(1)$        \\
$H_I^{(2)}(0)$ & $y'$      & $p$        & $n_l(1)$        \\
$H_I^{(1)}(1)$ & $y'$      & $p$        & $n_l(1)$        \\
$H_I^{(2)}(1)$ & $x'$      & $q$        & $n_l(1)$        \\
$H_I^{(1)}(2)$ & $x'$      & $q$        & $n_l(3)$        \\
$H_I^{(2)}(2)$ & $y'$      & $p$        & $n_l(3)$        \\
\end{tabular}
\caption{SVD-control qubit and constrained bosonic representation used for each hopping subterm in the production H2-2 circuit.}
\label{tab:choice}
\end{table}
The SVD orientation was chosen together with the remaining circuit structure and subsequent compilation; we do not interpret the isolated-subterm comparison above as establishing a globally optimal orientation for the full Trotter circuit.

We emphasize that the Gray-code gate counts in Table~\ref{tab:graycomp} are diagnostic quantities for comparing equivalent representations. They do not correspond to the production synthesis used for the hardware circuits.

\subsection{Inter-subterm optimization and compilation}
The hopping terms are further ordered to enable cancellation of compatible ladder-operation subcircuits between $H_I^{(1)}(2)$ and $H_I^{(2)}(1)$, $H_I^{(1)}(1)$ and $H_I^{(2)}(2)$.
This optimization is separate from both the SVD orientation and the choice of bosonic representation.

The analytically simplified Trotter circuit is subsequently compiled together with the initial-state preparation using pytket for the H2-2 processor. Denoting the preparation circuit by $P$, the uncompiled single-step evolution by $U_T$, and the compiled implementation of $ U_T\circ P$ by $C$, we exploit the fact that $P$ consists only of Pauli-$X$ gates and therefore satisfies $P^2=I$. An $N$-step circuit can then be assembled as
\begin{align}
(C\circ P)^N \circ P
\end{align}
where $\circ$ denotes sequential circuit composition in execution order. This construction is logically equivalent to $U_T^N \circ P$, while retaining the compiler-level simplifications obtained by jointly compiling $P$ and $U_T$. The resulting production circuit contains 226 native ZZPhase gates per Trotter step.

\section{Other error mitigation methods considered}\label{app:other-qem}
In this appendix, we list other error mitigation methods we considered but did not adopt in the experiment.
\begin{itemize}
    \item Readout error mitigation~\cite{Nachman:2019zfw,Hicks:2020sbj,Hicks:2021uvm} was not employed because the readout error of the Quantinuum hardware is comparable to the error of a single two-qubit gate~\cite{PhysRevX.13.041052}, while the circuits considered in this work contain hundreds of two-qubit gates. Therefore, the total error is expected to be dominated by accumulated gate errors rather than readout errors.
    \item Dynamical decoupling (DD) is a widely used error-suppression technique for reducing decoherence during idle periods~\cite{PhysRevLett.82.2417,PhysRevA.79.062324}. We evaluated DD in the present circuits but found that it provided negligible improvement for the physical observables studied here.
    \item We also considered other error-mitigation techniques, including zero-noise extrapolation (ZNE)~\cite{He:2020udd,Pascuzzi:2021mhw} and probabilistic error cancellation (PEC)~\cite{PhysRevLett.119.180509,PhysRevX.8.031027}. Although their implementations differ, both approaches generally incur substantial experimental and sampling overheads: ZNE requires repeated circuit executions at multiple effective noise levels, while PEC requires noise characterization and quasiprobability sampling. Given the limited hardware budget available for this study, we did not employ these approaches.
\end{itemize}

\section{Estimation of effective depolarizing-noise parameters}
\label{app:depnoisestudy}
For depolarizing error mitigation, it is necessary to estimate the depolarizing noise probability $p$ independently of the observables of interest. 
We consider three different methods and evaluate the reliability of each method using the H2-Emulator.

The first method determines a global depolarizing probability using an auxiliary electric-energy observable $H_E(3)$ constructed from the right-boundary flux. Although $r=3$ does not correspond to a dynamical lattice link, this quantity is fixed by the boundary condition and therefore has a known ideal expectation value.
The depolarizing noise probability $p$ is then estimated from the deviation of the measured value from a known ideal expectation value.

The second method executes circuits in which all $n_i(r)$, $n_o(r)$, and $n_l(r)$ in the initial state are set to zero, and estimates the effective noise parameter for each observable. 
In this initial state, the measurement outcome is always zero; thus, the ideal expectation value in the absence of noise can be easily determined, in the same manner as in the first method.

The third method starts from the nominal initial state but inserts pairs consisting of a Trotter-step circuit and its inverse, allowing the effective noise parameter for each observable to be estimated every two steps. 
Since each inserted pair acts as an identity operation, the final state should be the same as the initial state, and thus the ideal expectation value can be readily calculated. 
For a stationary depolarizing process, the survival factor $1-p$ composes multiplicatively with circuit depth. Over the limited depth range considered here, we approximate the depth dependence of the effective parameter linearly and use this empirical fit to estimate its value at each circuit depth.
Another issue with this method is that the noise parameter for link 2 cannot be estimated because $H_{E}(r=2,t=0) = 2.0$ is equal to the coefficient of the identity, $o_{\text{tr}}$.
\begin{figure}[hpb]
    \centering
    \subfloat[Method 1: Global noise probability]{\includegraphics[width=0.95\linewidth]{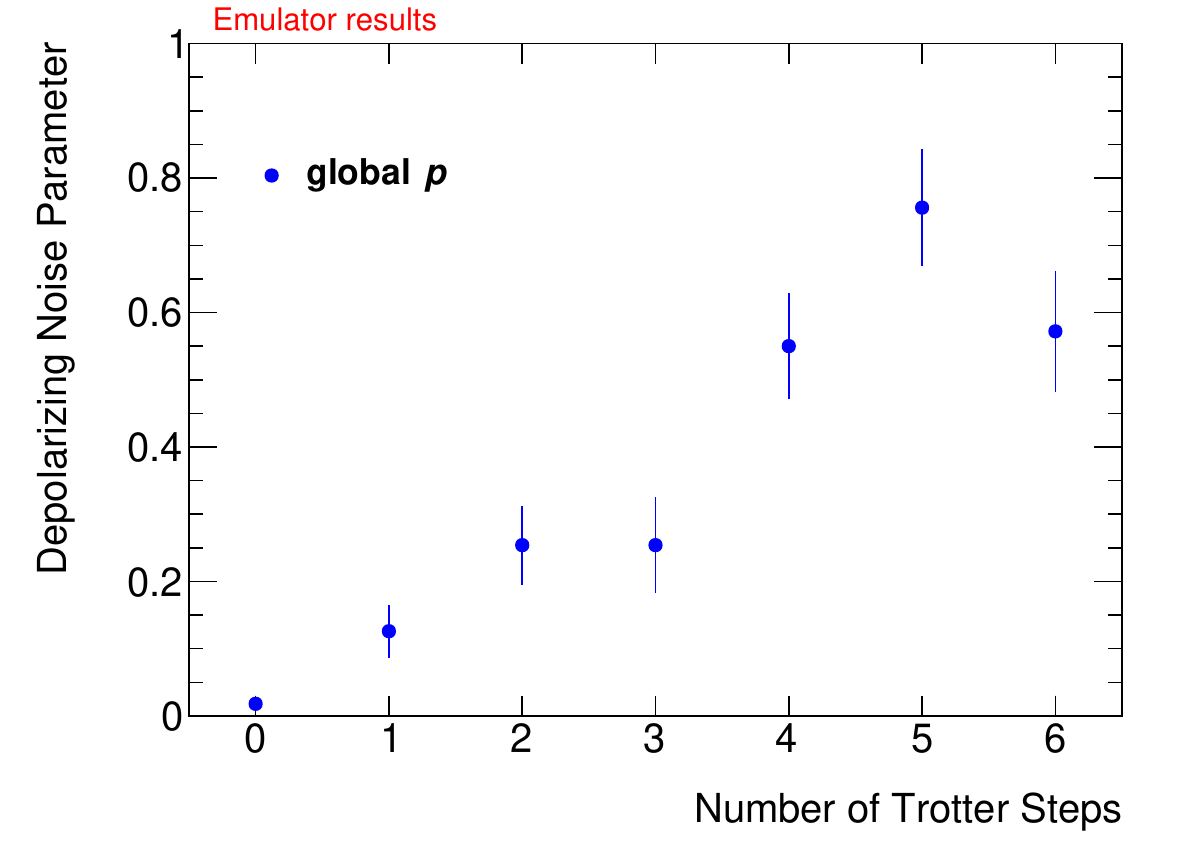}}
    \label{fig:p_method1}

    \subfloat[Method 2: Local noise parameter (zero-state initialization)]{
    \includegraphics[width=0.95\linewidth]{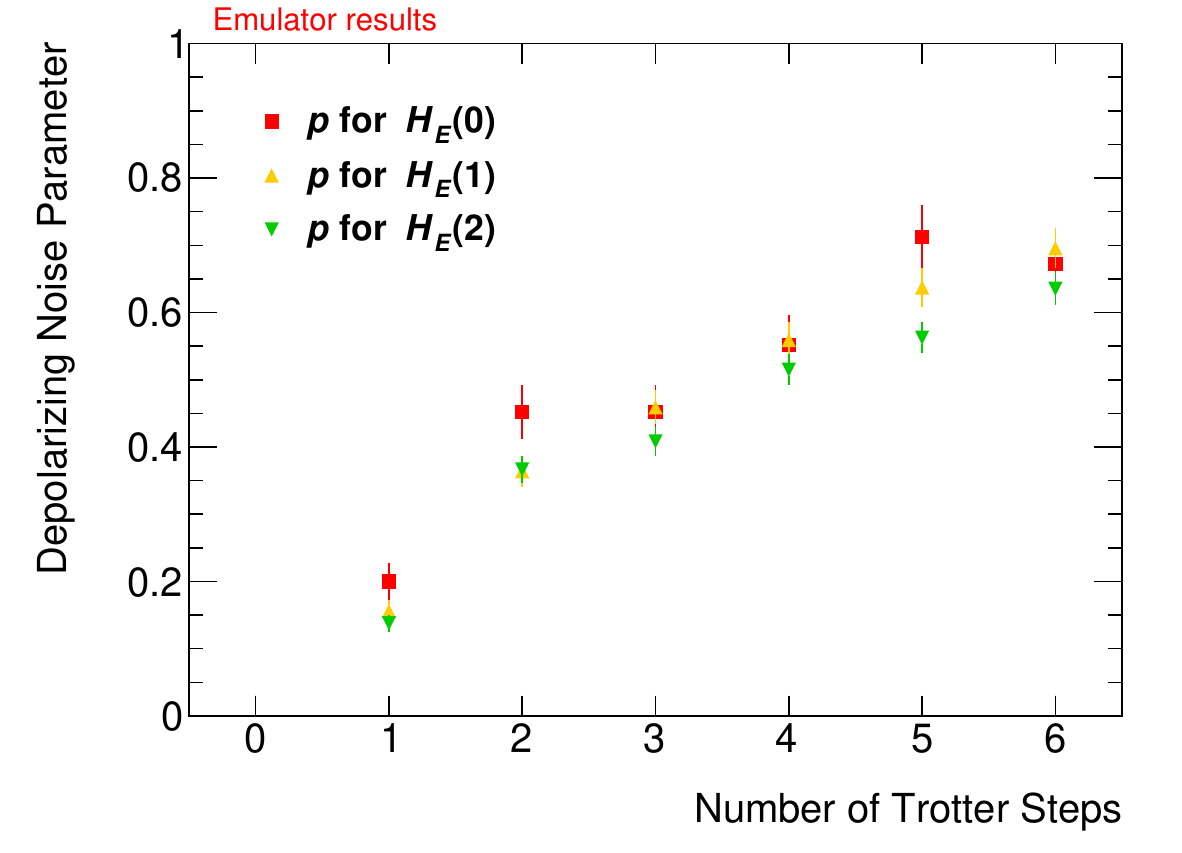}} \\
    \label{fig:p_method2}

    \subfloat[Method 3: Local noise parameter (Trotter step inverse pairs)]{
    \includegraphics[width=0.95\linewidth]{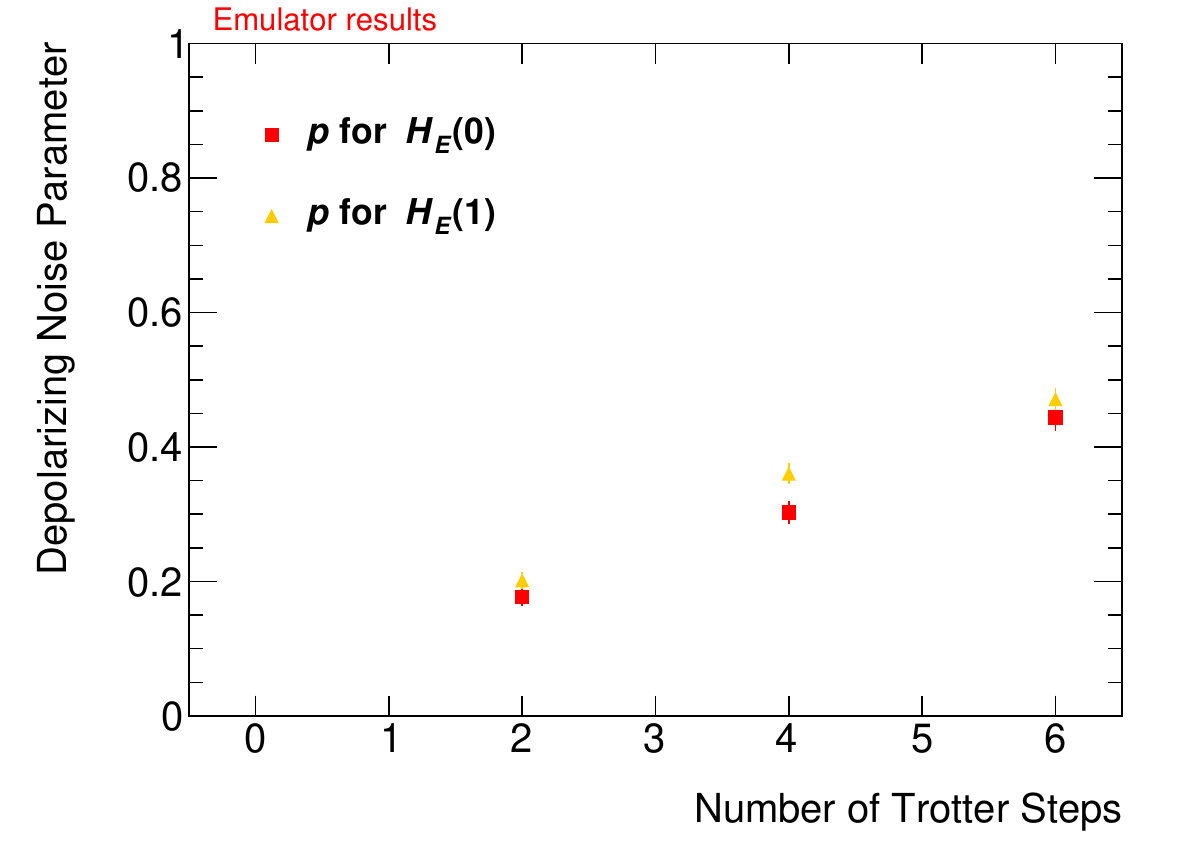}}
    \label{fig:p_method3}
    \caption{Estimated depolarizing noise probabilities in three different approaches. }
    \label{fig:noiseprob}
\end{figure}

\begin{figure}
    \centering
\subfloat[Differences between the ideal expectation values calculated by a noiseless simulator and the error-mitigated expectation values.]{
    \includegraphics[width=0.95\linewidth]{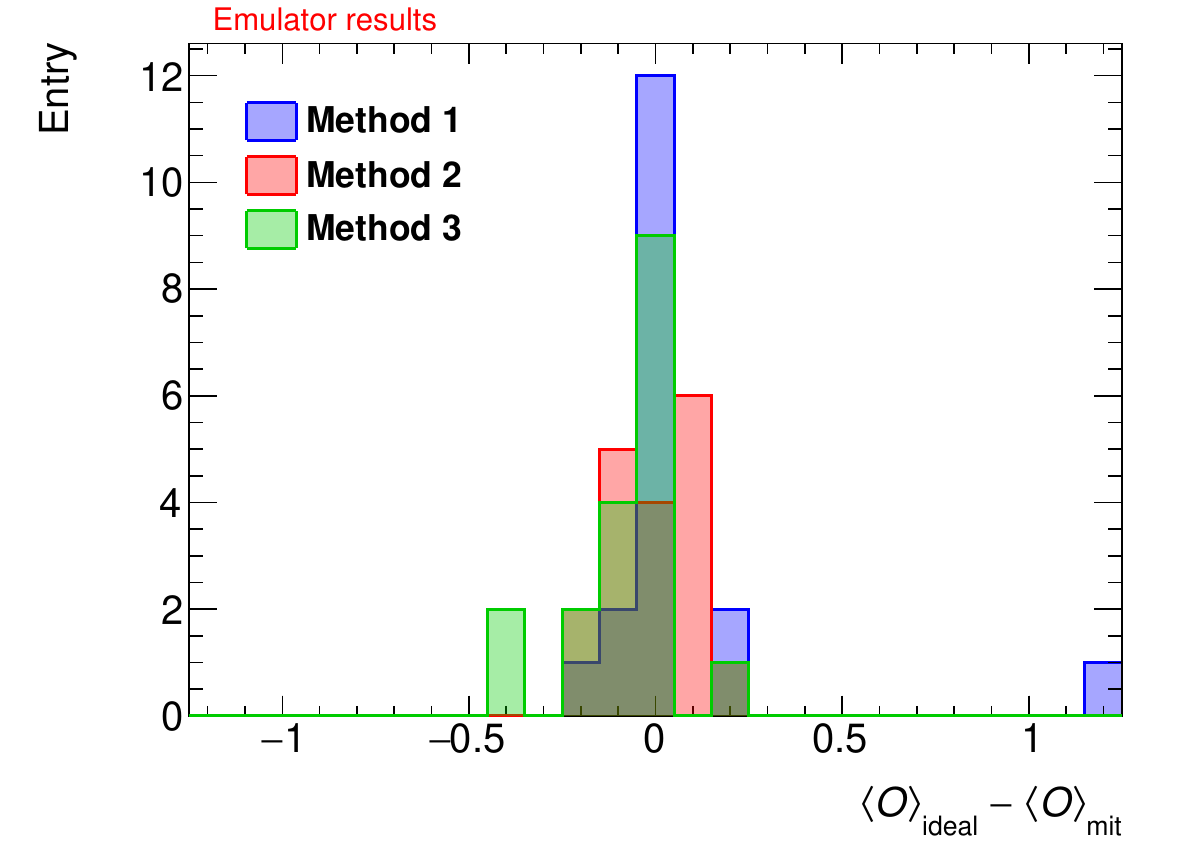}}
    \label{fig:comp_diff}

\subfloat[Significances, defined as the differences divided by the propagated statistical uncertainties $\delta\braket{O}_{\text{mit}}$, including contributions from both the measured expectation values and the estimated noise parameters.]{
    \includegraphics[width=0.95\linewidth]{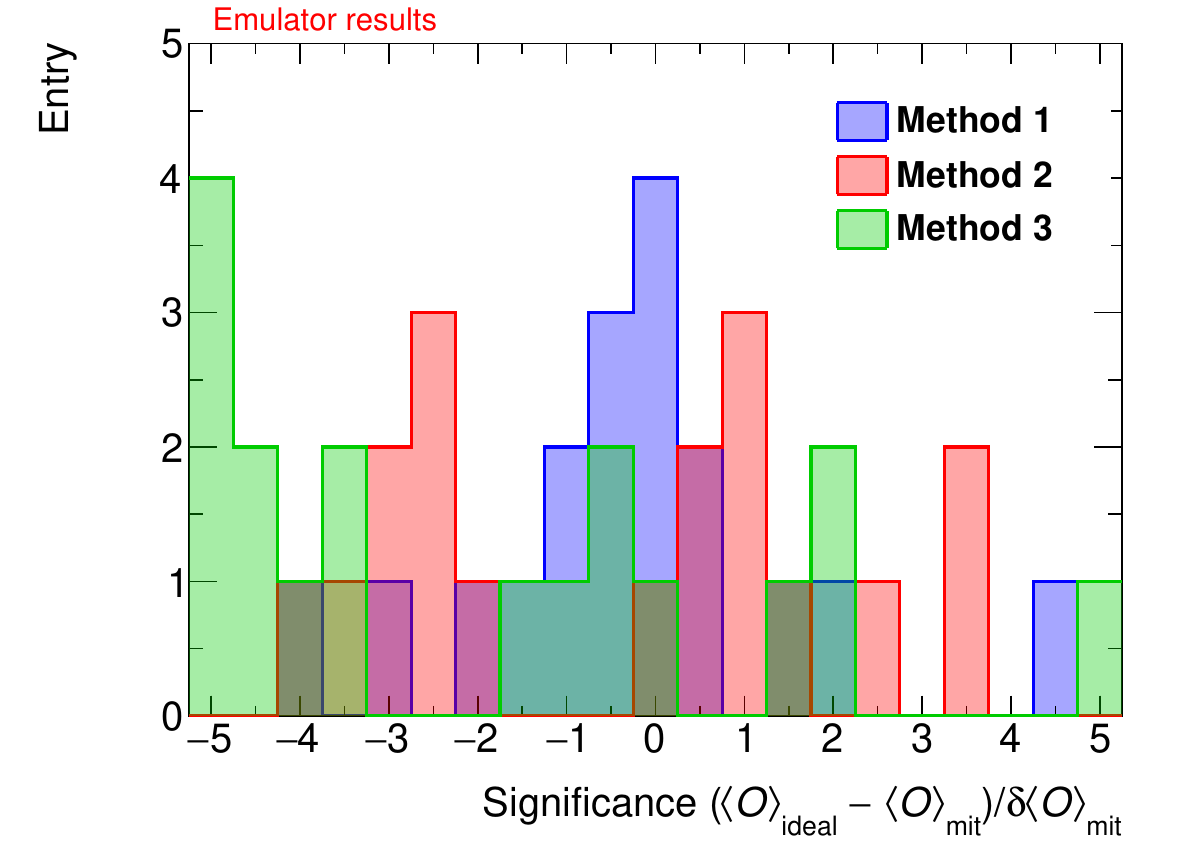}}
    \label{fig:comp_sig}
    \caption{Comparison of depolarizing noise parameter estimation methods.}
    \label{fig:comp_noiseprob}
\end{figure}

In all three methods, the ideal expectation values are known a priori and are independent of the observables of interest. 
Fig.~\ref{fig:noiseprob} shows the estimated noise probabilities. 
All three estimation methods yield effective noise parameters that generally increase with the number of Trotter steps, consistent with the accumulation of circuit noise with increasing depth. Moreover, the methods show broadly similar depth dependence, and none exhibits qualitatively anomalous behavior over the range studied here. These observations do not by themselves establish a depolarizing noise model, but indicate that the effective-parameter description is not obviously inconsistent with the emulator results.
Since there is no significant difference among the observables in Method 3, we use the parameter obtained for the link-0 in place of that for the link-2.
To evaluate the performance of each method, we compare the error-mitigated values with the ideal expectation values, as shown in Fig.~\ref{fig:comp_noiseprob}. 
Method 1 generally performs well, but it exhibits the largest discrepancy and also has the largest uncertainty in its estimated probability, which arises from the statistical errors of the noisy expectation values, $\delta\langle O \rangle_{\text{noisy}} = \sigma/\sqrt{n}$. 
Since the number of measurement shots is the same across all methods, this indicates that the observed distribution of the expectation values in Method 1 has the largest variance.
Although the discrepancy in Method 3 is relatively small, its statistical significance is high. 
This indicates that the deviation in Method 3 cannot be explained solely by statistical fluctuations. 
Therefore, we conclude that Method 2 is the most reliable approach, and we employ Method 2 in this paper.

\section{Exploratory study of Trotterization effects}
\label{app:trotterstudy}
In the main text, we benchmark the hardware results against the corresponding noiseless first-order Trotterized-circuit results rather than the exact time evolution, thereby separating hardware-induced deviations from Trotterization effects.
In this appendix, we first illustrate the size of the Trotterization error by comparing the exact time evolution with noiseless first-order Trotterized calculations.
We then investigate Trotter-error mitigation using circuits with varying numbers of Trotter steps at fixed total evolution times, with the aim of obtaining a more accurate approximation to the exact time evolution.
However, this mitigation approach proves ineffective in the presence of circuit noise, while the single-step second-order Trotterized circuit provides a more accurate approximation to the exact time evolution over the regime studied.
Accordingly, this appendix presents the emulator-based study of Trotter-error mitigation together with the results obtained using the single-step second-order circuit.

\begin{figure}
    \centering
    \includegraphics[width=0.95\linewidth]{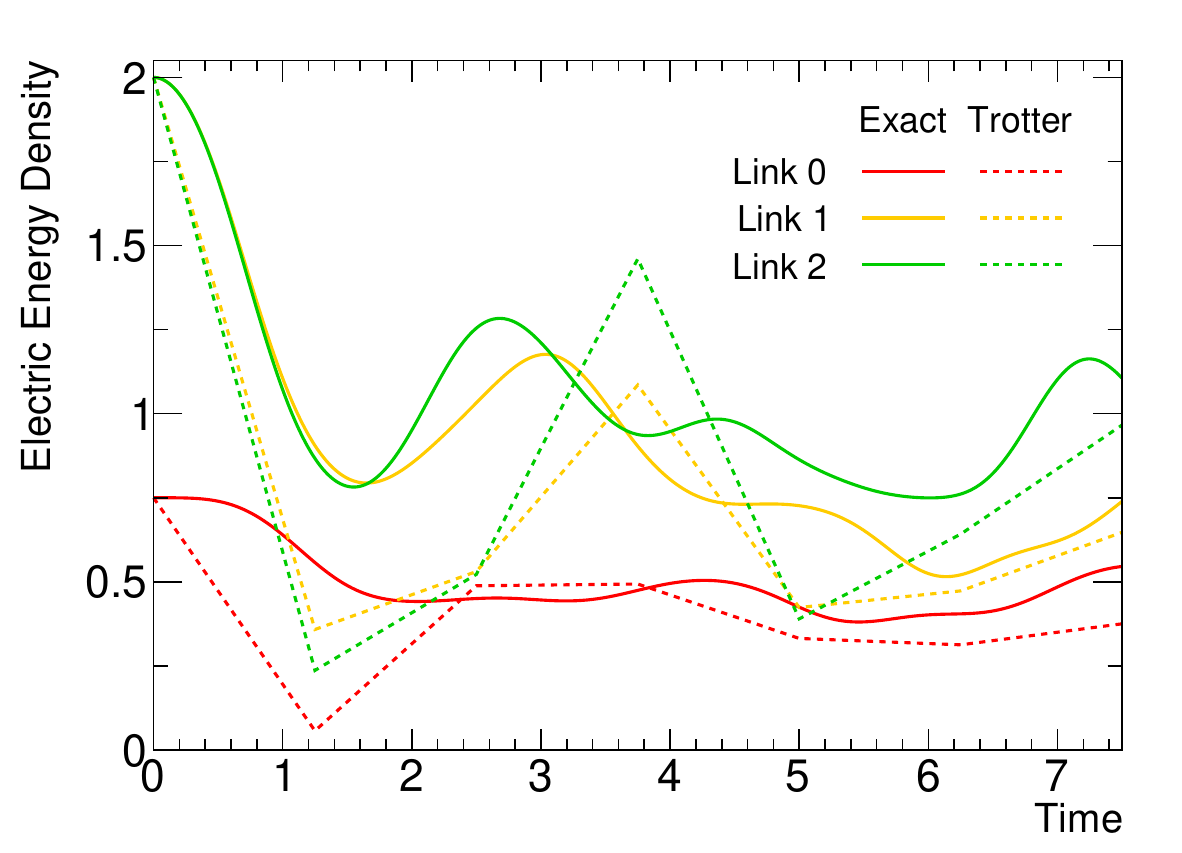}
    \caption{Comparison of the exact time evolution and the noiseless first-order Trotterized evolution with $\Delta t=1.25$ used in the main hardware study. Solid and dashed lines represent the exact and Trotterized results, respectively, for the electric-energy densities on links 0, 1, and 2.}
    \label{fig:compexact}
\end{figure}

Fig.~\ref{fig:compexact} compares the exact time evolution with the noiseless first-order Trotterized evolution at $\Delta t=1.25$, as used in the main hardware study, over the same evolution-time range $0 \leq t \leq 7.5$. The exact evolution exhibits nontrivial oscillatory behavior in the electric-energy densities on all three links, and this qualitative time dependence is also retained in the Trotterized results. However, sizable quantitative deviations are visible, including shifts in the positions and amplitudes of the extrema. These differences represent the Trotterization effects that are separated from hardware-induced deviations in the main-text analysis.

\begin{figure}[t]
    \centering
\subfloat[$T=0.2$ on link-2]{
    \includegraphics[width=0.95\linewidth]{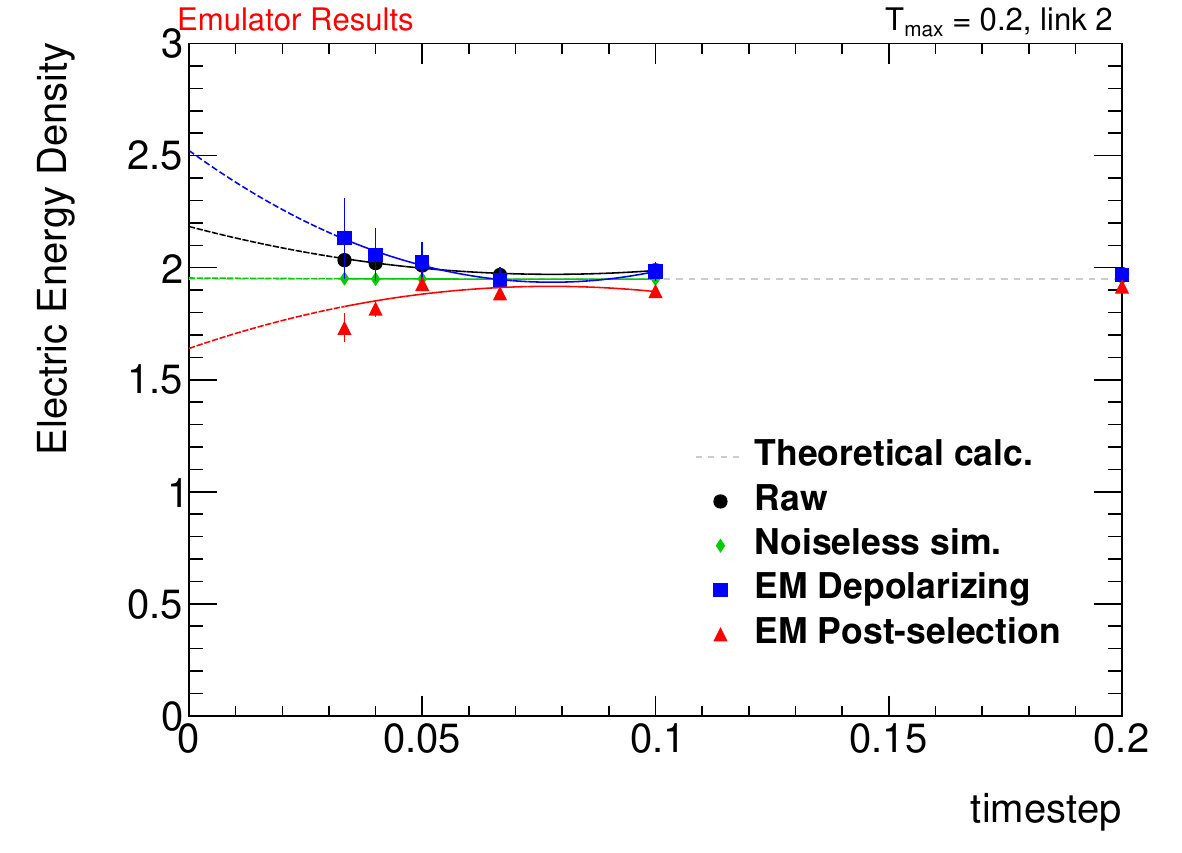}}
    
\subfloat[$T=1.0$ on link-2]{
    \includegraphics[width=0.95\linewidth]{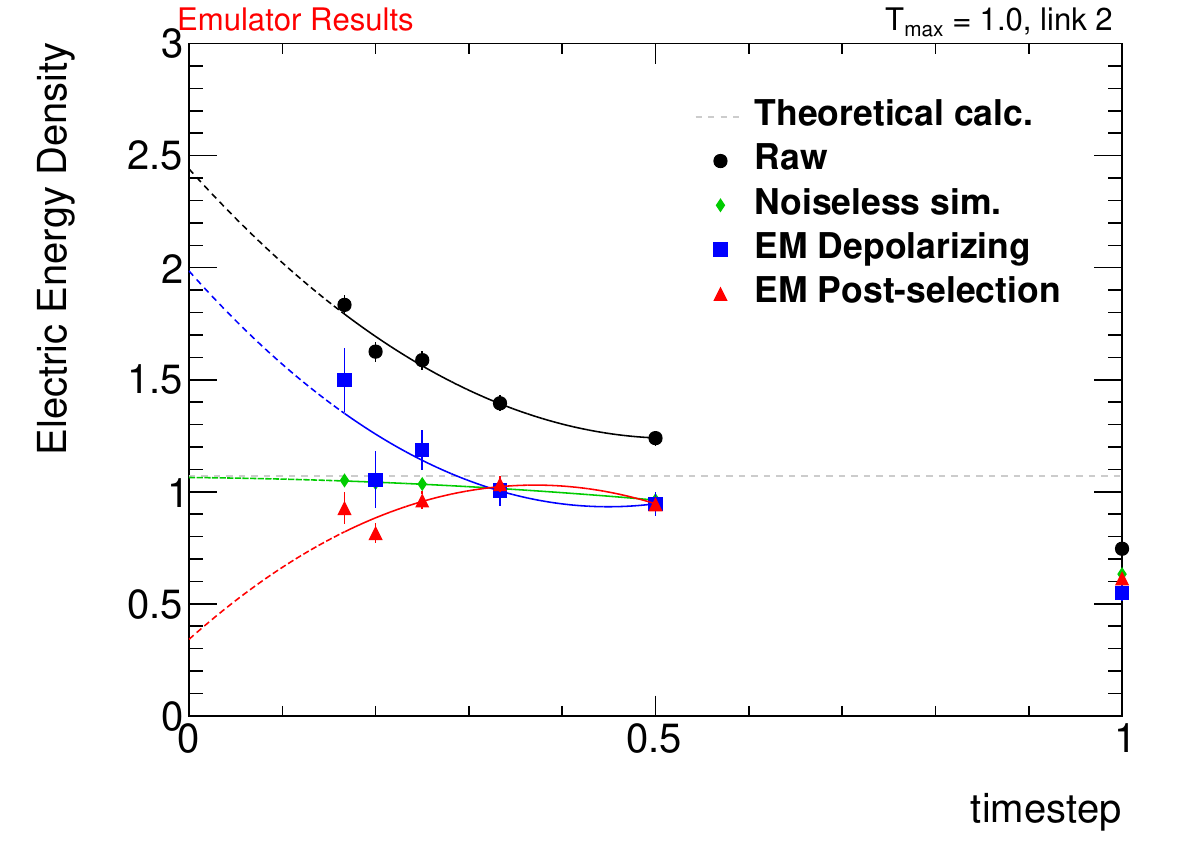}}
    \caption{Extrapolation for Trotter error mitigation. The horizontal and vertical axes represent timestep and electric energy density, respectively. The dotted gray line indicates the exact time evolution, while the markers denote the results from the Quantinuum emulator: black circles for raw observed values, green diamonds for noiseless values, blue squares for depolarizing error-mitigated values, and red triangles for error-mitigated values using post-selection. The five leftmost points are used for the quadratic fitting. The solid and dashed lines in each color represent the fitted functions inside and outside the fitting range, respectively.}
    \label{fig:TEM}
\end{figure}
Motivated by these finite-timestep deviations, we investigated whether the Trotterization error could be reduced by extrapolating to the zero-timestep limit. To this end, we execute quantum circuits with varying numbers of Trotter steps at a fixed evolution time, $T$, allowing us to obtain results calculated across different timesteps. 
We then estimate the expectation value at a timestep of 0 (i.e., the expectation value without Trotter errors) by extrapolating with a quadratic function. 
We apply this extrapolation procedure to the observed values both with and without error mitigation. 
The error mitigation strategy is identical to that used in the main text. 
Fig.~\ref{fig:TEM} shows the fitting results for $T=0.2$ and $T=1.0$ on link-2. 
We exclude the 1-step data point from the fitting, as our preliminary emulator study confirmed that its exclusion yields more precise estimation results. 
The vertical intercept of the fit function represents the target value extrapolated to the ideal zero-Trotter-error limit. 
In the very low-$T$ regime ($T=0.2$), the Trotter error is sufficiently small. 
However, at $T=1.0$, a clear discrepancy emerges due to the Trotter error. 
While the noiseless results demonstrate the validity of the extrapolation procedure itself, the method proves ineffective for noisy data. 
Because the most critical left-most point contains the largest circuit noise, the estimated values strongly depend on the fluctuations of the observed values. 
Therefore, we conclude that we lack sufficient precision to effectively apply this Trotter error mitigation method.
\begin{figure}[t]
    \centering
    \includegraphics[width=1.0\linewidth]{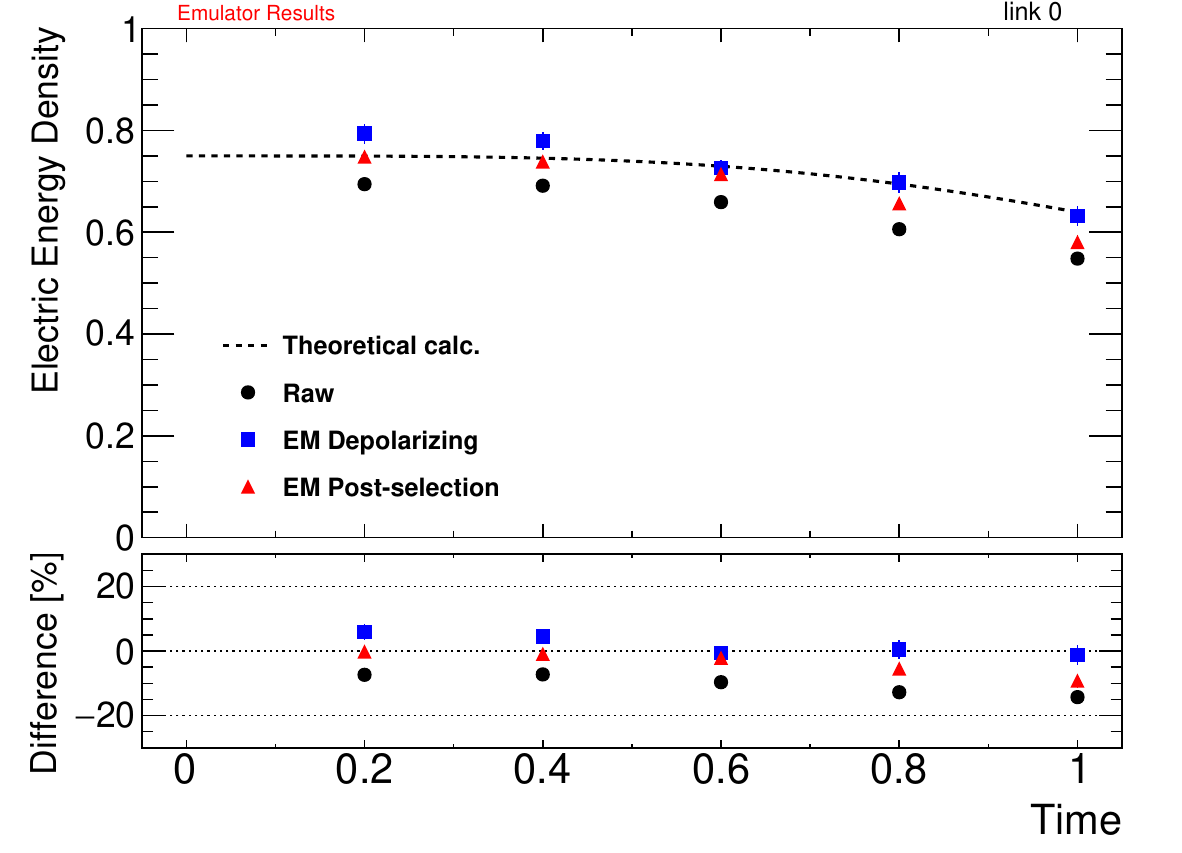}  \\
    \includegraphics[width=1.0\linewidth]{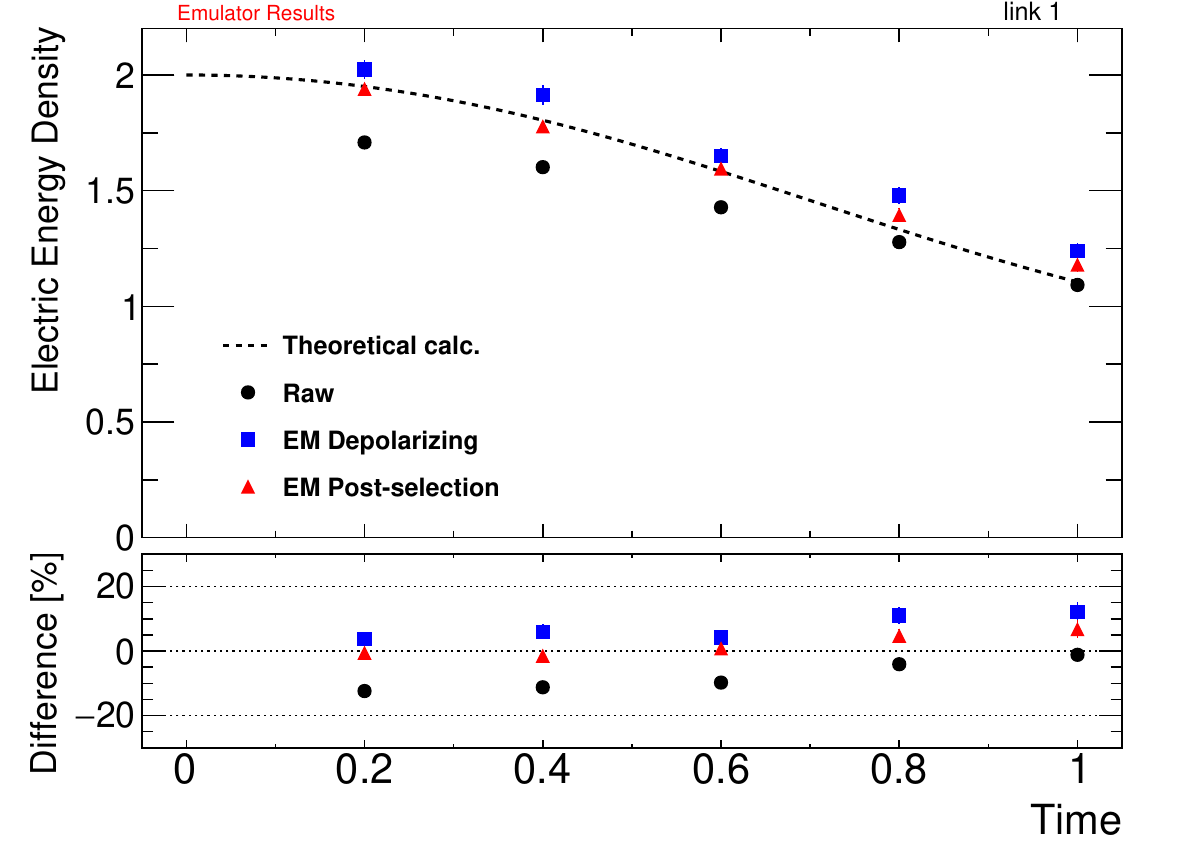} \\
    \includegraphics[width=1.0\linewidth]{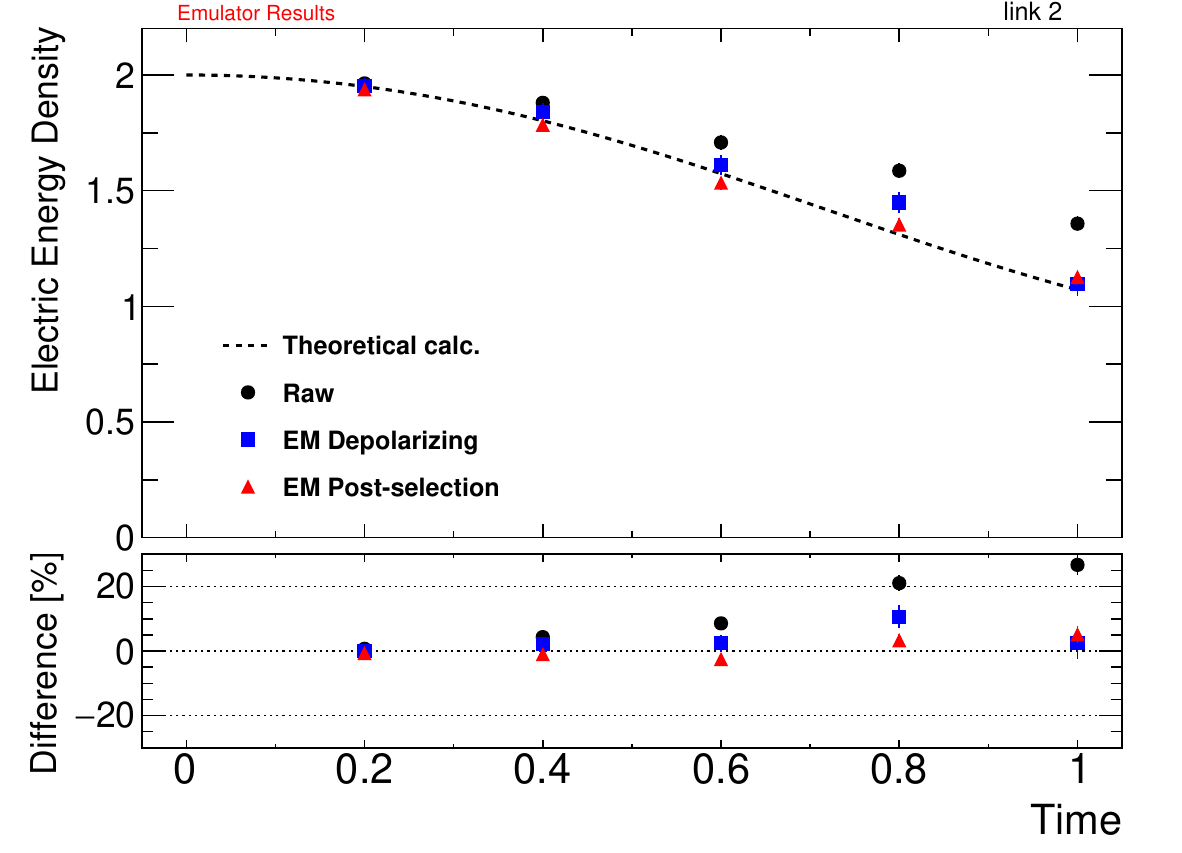}
    \caption{Electric energy density at each lattice link calculated using a single-step second-order Trotterized circuit. The horizontal and vertical axes represent dimensionless time and electric energy density, respectively. The dotted line indicates the exact time evolution, while the markers denote the results from the Quantinuum emulator: black circles for raw observed values, blue squares for depolarizing error-mitigated values, and red triangles for error-mitigated values using post-selection.}
    \label{fig:elec-2ndorder-QEM}
\end{figure}

Another approach to achieve higher precision is to employ second-order Trotterization. 
The Trotter error scales with the timestep; at $t=1.0$, it is approximately 8--10\% and 0--3\% for the 1- and 2-step circuits, respectively in second-order Trotterization. 
There is a trade-off between Trotter error and circuit noise. 
The constructed circuit consists of 381 two-qubit (\texttt{ZZPhase}) gates per Trotter step, which is about 70\% higher than that of the first-order circuit. 
Fig.~\ref{fig:elec-2ndorder-QEM} shows the electric energy density calculated using a single-step second-order Trotterized circuit up to $T=1.0$. 
The deviation from the exact time evolution is reduced by error mitigation. 
We also show the relative difference from the theoretical values, calculated as $(\text{Data} - \text{Theory})/\text{Theory}$. 
Given that the maximum relative difference is 12\% using depolarizing error mitigation and 9\% using post-selection, the hardware runs as presented in the main text achieved an accuracy on the same order of magnitude as the Trotter error itself.

\bibliography{refs}
\end{document}